# Non-parametric Ensemble Empirical Mode Decomposition for extracting weak features to identify bearing defects

Anil Kumar[1*], Y. Berrouche[2], Radoslaw Zimroz[3], Govind Vashishtha[4], Sumika Chauhan[5], C.P. Gandhi[6], Hesheng Tang[1], Jiawei Xiang[1]

[1]College of Mechanical and Electrical Engineering, Wenzhou University, Wenzhou, 325 035, China.

[2]LIS laboratory, Department of Electronics, Faculty of Technology, Ferhat Abbes Sétif 1 University, Algeria

[3]Wroclaw University of Science and Technology, Wroclaw 710072, Poland

[4]Sant Longowal Institute of Engineering and Technology Longowal, 148 106, India

[5]Department of Computer Science, National Institute of Technology Delhi, New Delhi 110 036, India

[6]Rayat Bahra University, Mohali, 140 104, India

*Corresponding Author email: anil_taneja86@yahoo.com

**Abstract:** A non-parametric complementary ensemble empirical mode decomposition (NPCEEMD) is proposed for identifying bearing defects using weak features. NPCEEMD is non-parametric because, unlike existing decomposition methods such as ensemble empirical mode decomposition, it does not require defining the ideal SNR of noise and the number of ensembles, every time while processing the signals. The simulation results show that mode mixing in NPCEEMD is less than the existing decomposition methods. After conducting in-depth simulation analysis, the proposed method is applied to experimental data. The proposed NPCEEMD method works in following steps. First raw signal is obtained. Second, the obtained signal is decomposed. Then, the mutual information (MI) of the raw signal with NPCEEMD-generated IMFs is computed. Further IMFs with MI above 0.1 are selected and combined to form a resulting signal. Finally, envelope spectrum of resulting signal is computed to confirm the presence of defect.

## 1. Introduction

Rotating components such as bearings, gears, etc., are the core elements being used in modern manufacturing units and hence should be reliable as these components are prone to defects. The induced fault in rotating components may increase the risk of failure which can affect the production yields resulting the increase in the production cost [1–3]. In order to avoid these sensitive issues of rotating components, regular monitoring is essential to prevent sudden failure. Vibration monitoring can be used to aid in the early detection of faults [4–7]. A vibration signal is extremely complicated source of rotating components and vibration-based diagnosis approach merely focus on the evaluation of the system's critical components.

Because the rotating machinery has many sources of vibration, the signal exhibits many characteristics of multiple frequencies [8–10]. A reliable fault detection methodology must include the extraction of defect features, which can be effectively performed through signal decomposition and filtering approach. To tackle with the signals having many frequency components, the eminent researchers have successfully developed adaptive methodologies [11,12] including empirical mode decomposition (EMD) and non-adaptive decomposition techniques based on wavelet transform [13–17].

The non-adaptive decomposition techniques are problematic to utilize because the decomposition level and wavelet are to be selected beforehand. Whereas EMD is an adaptive decomposition technique that attempts to decompose the original signal into several intrinsic mode functions (IMFs) [18,19]

Researchers have been taking a great interest in developing EMD based fault diagnosis method because of its numerous advantages. Yu and Cheng [20] employed energy entropy (calculated from signal processed by EMD), for the diagnosis of bearing defects using ANN. Li and Zhang [21] used Wigner-Ville distribution and EMD to identify various bearing defects. Li et al. [22] applied EMD approach. to detect abnormal clearance of a diesel engine. EMD has been proved to be a reliable and exceptional tool for identifying bearing faults But, it has several limitations that require kind attention. One of the inadequacies is the separation issue [23], in which EMD has been found inefficient in differentiating those components whose frequencies lie inside a band. Another drawback of EMD is that it is subjected to noise, which can be referred to as the problem of interruption [23]. As a result of which, it returns either a single IMF consisting of multiple scales or multiple IMFs containing a signal with the same scale. The ensemble empirical mode decomposition (EEMD), which is a variant of EMD, can be computed by adding white noise before decomposition [24,25]. The issue of mode mixing may be circumvented to some extent through the employment of technologically sophisticated EEMD. Although, EEMD was developed a decade ago, however, it has still been applying for decomposing the signal. Lu et al. [26] proposed the zero-shot intelligent fault diagnosis scheme and extracted statistical attributes of the decomposed signals from EEMD and finally obtained time and frequency characteristics. With the attributes of known conditions, the authors constructed the well-known Gaussian model for the purpose of classifying the attributes of unknown conditions. The authors have been able to achieve an identification accuracy of 86 % while diagnosing the faults through the proposed methodology. Zhou et al. [27] proposed a robust two-stage denoising method by hybridizing the EEMD and independent component analysis (ICA) by considering the fuzzy entropy discriminant as a threshold. The CNN based on the VGG structure has also been incorporated as a classifier in the proposed method. As traditional sparse representation approaches take more computational time, Wang et al. [28] incorporated the EEMD with the improved sparse representation approach and established orthogonal matching pursuit (adapOMP) algorithm and effectively extracted the harmonic components required for the filtered signal. The filtered signal is further decomposed through the EEMD having the kurtosis value greater than three. Thereafter, Hankel matrix transformation is carried out to construct the learning dictionary. However, EEMD has two serious drawbacks. First, EEMD requires optimum amplitude of white noise and ensemble size every time signal is to be processed, which is a very difficult task. Second, the issue of separation is not entirely addressed by the EEMD. Complementary ensemble empirical mode decomposition (CEEMD) [29] is another decomposition method in which white noise is added in pairs to the original data (i.e., one positive and one negative) to generate two sets of ensemble IMFs. Then, the final IMFs are obtained by averaging two sets of IMFs. However, CEEMD performs extremely poorly when the wrong magnitude of white noise is specified. This means that every time a signal needs to be processed, there is a need to define the noise's magnitude. Complete ensemble empirical mode decomposition with adaptive noise (CEEMDAN) [30] is a novel variant of the EEMD and can offer different noise realizations at each stage of signal decomposition. Despite of different noise realizations at each stage, CEEMDAN still face mode-mixing problem.

The non-parametric ensemble empirical mode decomposition (NPCEEMD) [31] is proposed as a solution to two problems, mode mixing and parametric dependent. The NPCEEMD methodology works on the principles, similar to CEEMD. In NPCEEMD, a fractional Gaussian noise is used in place of white Gaussian noise. The fractional noise has been found more abundant at high frequencies and thus making it suitable for dealing with the signal's high-frequency noise. Also, NPCEEMD is a non-parametric method as it does not require selecting the appropriate SNR and number of ensemble trials every time for each signal processing step. This makes the NPCEEMD a better choice for those situations under which processing time is very important and huge amount of data is to be handled.

In this study, we have successfully established a non-parametric complementary ensemble empirical mode decomposition (NPCEEMD) based methodology to extract weak defect features for the identification of bearing defects. NPCEEMD based methodology works in following steps. First raw signals are obtained. After obtaining the raw signals, NPCEEMD is used to decompose the vibration signals. Then mutual information (MI) between IMFs and raw signal is computed. The IMFs correlated with the raw signal are supposed to have a high amount of MI. IMFs having high MI with raw signal are selected and combined to form resulting IMF. Finally, envelope spectrum of the resulting IMF is computed to ensure the presence of bearing defect. The analysis results indicate that our proposed NPCEEMD-based methodology furnishes better results and is computational efficient compared to the existing decomposition techniques.

The author's major contribution in this study can be summarized as follows:

1) A non-parametric ensemble empirical mode decomposition (NPCEEMD) based methodology is proposed for the identification of defects in the rotary machinery which does not require selection of optimum parameters every time signal to be processed

2) A method based on mutual information is proposed to select the IMF having defect related information. The superiority of the proposed band selection method is demonstrated against the existing kurtosis-based band selection method.

Through, simulation and experimental analysis it has been showed the decomposition performance of the proposed non-parametric ensemble empirical mode decomposition (NPCEEMD) is better. The simulation results verify that the proposed method is non-parametric and easily suits every signal. Results show that non-parametric ensemble empirical mode decomposition (NPCEEMD) has better results than existing decomposition methods such as empirical mode decomposition (EMD), ensemble empirical mode decomposition (EEMD), complementary ensemble empirical mode decomposition (CEEMD), and complete ensemble empirical mode decomposition with adaptive noise (CEEMDAN). The organization of the rest of the paper is as follows: **Section 2** describes the theoretical aspects of the existing decomposition methodologies, needed for the subsequent development of the proposed study. In **Section 3**, the decomposition performance and non-parametric behaviour of the proposed non-parametric ensemble empirical mode decomposition (NPCEEMD) is evaluated and compared with the existing methodologies. The effect of Hurst exponent on the decomposition performance of the proposed NPCEEMD based methodology is examined in **Appendix A.1**. In **Appendix A.2**, the effect of ensemble number on the decomposition performance of the proposed as well as existing methods is investigated. **Section 4** describes the proposed non-parametric complementary ensemble empirical mode decomposition (NPCEEMD) based methodology to extract weak defect features for identifying bearing defects in rotary machinery. In **Section 5**, the applicability of the proposed NPCEEMD based methodology has been represented whereas **Section 6** summarizes the concrete conclusion drawn from the proposed work.

## 2. Theory

This section is devoted to describe the theoretical concept of various signal processing methods , needed for the proposed study.

### 2.1 Empirical mode decomposition (EMD)

The EMD algorithm decomposes an input non-linear and non-stationary signal into a finite number of IMFs (intrinsic mode functions) components by using a sifting process. In EMD decomposition, a signal $x(t)$ is a superposition of a finite number of IMFs and a low-order polynomial called as residue and denoted by $R(t)$[32]:

$$x(t) = \sum_{i=1}^{N} IMF_i(t) + R(t) \qquad (1)$$

where $N$ is the number of IMFs obtained during the iteration process, $IMF_i(t)$ is the i$^{th}$ IMF generated out of the signal $x(t)$.

The major problem with EMD is its accuracy which may be compromised by a strong mode mixing problem.

### 2.2 Ensemble Empirical Mode Decomposition (EEMD)

For a given number of ensemble trials (Ne), the EEMD algorithm is a noise-assisted data processing method in which a finite amount of white noise is added to an input signal. Then, the resulting signal is decomposed by the EMD algorithm to generate the IMFs. By repeating the steps Ne times, the true ensemble IMFs sets are defined in order to calculate the mean of all ensemble trials obtained from the IMFs of same order [33].

In fact, before applying the EEMD algorithm, the user must fix the appropriate amplitude of white noise and number of ensemble trials (Ne) before signal processing. Although, the EEMD method has successfully resolved the mode mixing problem, but the EEMD algorithm is a parametric method in which a large number of ensemble trials is required to cancel the effect of added white noise and this may increase the computation time. Likewise, the selection of parameters differs from one signal to another which makes the EEMD algorithm a non-standard method.??

### 2.3 Complementary ensemble empirical mode decomposition (CEEMD)

The CEEMD algorithm is an improvement of the EEMD. In the CEEMD algorithm, a white gaussian noise $(n)$ is added and subtracted from the input signal $(x)$ to generate two signals ($r_1$ and $r_2$)

$$r_1 = x + n \qquad (2)$$

$$r_2 = x - n \qquad (3)$$

The resulting two signals **(Eqs. 2-3)** are decomposed by the EEMD algorithm to generate two ensembles of IMFs. The true IMFs are then obtained by averaging the IMFs of same order [29]. However, the IMFs obtained by the CEEMD algorithm are less affected by noise than in the EEMD method. But, before applying the CEEMD algorithm, the user needs to select the desired parameters for every input signal which makes this algorithm a non-standard method.

## 2.4 Non-parametric ensemble empirical mode decomposition (NPCEEMD)

The NPCEEMD algorithm also uses the similar approach as adopted by the CEEMD algorithm. But, in the NPCEEMD algorithm, the fractional gaussian noise (FGN) is replaced by the Gaussian white noise [31]. The fractional Gaussian Noise (FGN) is a stationary Gaussian-centred random process, generated periodically by sampling the fractional Brownian motion phenomenon $(B_H(t))$. The FGN is richer in high frequency which makes it suitable to deal with high-frequency noise in the signal. As a result, the signal of interest can easily be extracted.

The computation of the first difference is defined by [31]:

$$x(n) = B_H(n) - B_H(n-1) \qquad (4)$$

The autocorrelation of FGN of lag $k$ is given by:

$$R_x(k) = \frac{\sigma^2}{2}(|k-1|^{2H} - 2|k|^{2H} + |k+1|^{2H}) \qquad (5)$$

The FGN is characterized by the Hurst exponent H and its variance $\sigma^2$, where $0 < H < 1$.

### 2.4.1 Fractional Gaussian noise process (FGP)

**Definition:** A stochastic process $X_t; t \geq 0$ becomes a Gaussian process if any finite linear combination of $x(t); t \geq 0$ is Gaussian. A Gaussian process can be denoted by $G_t^H; t \geq 0$ where $H$ is the Hurst exponent or index. Let $G_t^H, G_s^H; t \geq 0$ be any two Gaussian processes. A Gaussian process is called as fractional Gaussian noise process (FGP) if its mean function is $E(G_t^H) = \mu$ and covariance function is given by **Eq.6** [34].

$$COV(G_t^H, G_s^H) = \frac{\sigma^2}{2}(|t-s+1|^{2H} + |t-s-1|^{2H} - 2|t-s|^{2H}) \qquad (6)$$

where $\mu \in R, \sigma \in R^+, H \in (0,1), s, t \in 0, \infty)$. The value of $H$ can be utilized to determine the following.

(i) If $H = 0.5$, then the fractional Gaussian process is the Brownian.

(ii) (ii) If $H < 0.5$, then the increments $(G_{t+1}^H - G_t^H)$ of the FGP are negatively correlated and the process exhibits the short-range dependence.

(iii) (iii) If $H > 0.5$, then the increments $(G_{t+1}^H - G_t^H)$ of the FGP are positively correlated and the process exhibits the long-range dependence.

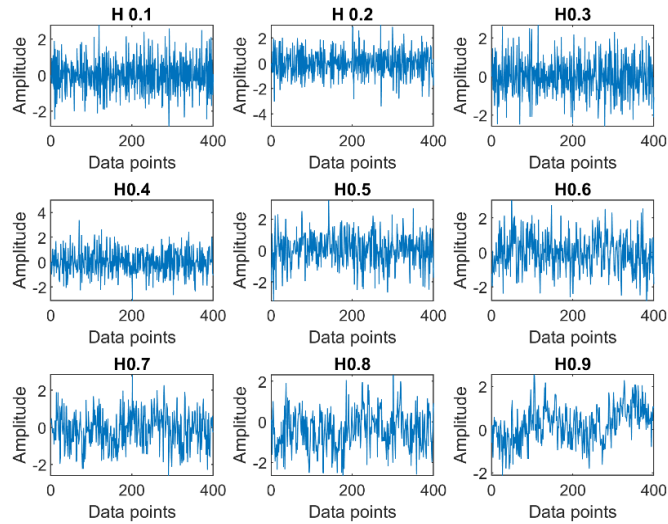

***Fig. 1.** Fractional gaussian noise of length 400 and different Hurst exponent*

Fig. 1 shows a fractional Gaussian noise of length 400 and with standard deviation 3. The value of Hurst exponent between 0.5 and 1 indicates a time series with long-term positive autocorrelation which means a high value will likely be followed by another high value and future values will likewise tend to be high. The value of Hurst exponent between 0 and 0.5 suggests a time series with long-term switches between high and low values in adjacent pairs. This means that a single high value will probably be followed by a low value, and the value after that, will likely be high. The value $H = 0.5$ may represent an uncorrelated series, but it can be applied to those series where autocorrelations at tiny time delays can be positive or negative, but fade exponentially to zero.

To demonstrate the effect of the Hurst exponent on the NPCEEMD performance, we need to compute the IMFs using different values of $H$. The IMF's using the values of $H$ between 0.1 to 0.9, are shown in **Figs. A1 to A9** in **Appendix A.1**. From these figs, we can conclude that whenever $H$ increases in value, the mode mixing becomes more prominent. Therefore H=0.1 is suitable for decomposing the signal.

## 2.5 Mutual information using the nearest neighbour approach

The concept of mutual information (MI) is an effective way to figure out how the two data sets are related. The mutual information between continuous variables using the nearest neighbour approach can be computed by the following procedure [35,36].

**Step 1:** First initialize the input parameters such as data and the number of nearest neighbors.

**Step 2**: Calculate the size of input data $x$.

**Step 3:** For the $i^{th}$ point, use the Chebyshev distance formula to find the nearest neighbors. Let us say this distance is $d_i$. The Chebyshev distance between two vectors or points $x$ and $y$, with standard coordinates $x_i$ and $y_i$, is given by max ($|x_i - y_i|$). The reason behind using the Chebyshev distance formula is that it gives the maximum distance between the two points along any coordinate direction, which is essential to find the nearest neighbors.

**Step 4:** Locate the index of data $x$ that are $k+1$ nearest neighbor from $i^{th}$ point of data $y$. Let us say that index is $N_{x_i}$.

**Step 5:** Investigate the $m_i$ points which are within the distance range as computed in Step 3.

**Step 6:** Compute the i$^{th}$ mutual information ($I_i$) by using Eq. 7

$$I_i = \psi(N) - \psi(N_{x_i}) + \psi(k) - \psi(m_i) \qquad (7)$$

where $N$ is the size of the dataset, $k$ is the number of nearest neighbors and $\psi$ is the digamma function,

**Step 7***:* Calculate the complete mutual information (MI)(Eq.8) by averaging the $I_i$th mutual information (Eq.7)

$$MI = \psi(N) - \langle \psi(N_{x_i}) \rangle + \psi(k) - \langle \psi(m_i) \rangle \qquad (8)$$

We now, equally well, proceed to demonstrate the decomposition performance of the proposed NPCEEMD method along with its comparison with the existing methods.

## 3. Evaluation of decomposition performance using simulation:

In this section, the simulation analysis is carried out to demonstrate that the decomposition performance of the proposed NPCEEMD method is better than the existing EMD, EEMD, CEEMD, and CEEMDAN-based methods. In essence, our goal is also to establish that our proposed NPCEEMD method is non-parametric.

For the simulation analysis, pure impulse and low-frequency signals are created separately and then combined. The IMFs have also been calculated using EMD, EEMD, and CEEMD for the purpose of highlighting the mode mixing issue of EEMD and how NPCEEMD can overcome the mode missing problem with minimal computational cost. Fig. 2(a) is the low-frequency signal generated using **Eq. 9**. The signal shown in **Fig. 2(b)** is an impulsive signal produced by **Eq. 10**. The combined signal can be obtained by simply adding the **Eqs. 9-10**, is shown in **Fig. 2(c).**

First, EMD is applied to the signal shown in **Fig. 2(c).** The resulting IMFs generated using the EMD method are shown in **Fig. 3**. **Fig. 3** indicates that the low-frequency signal has been recovered, but impulses cannot be obtained due to mode mixing.

$$x = 10 \times sin(2 \times 20 \times \pi \times t) \tag{9}$$

$$y_{imp} = 100 \times sin(2 \times 1400 \times \pi \times t) \times exp(-900t_2) \tag{10}$$

Here $t$ is the time vector of length 32768, t = [1:1:32768]/fs; fs is the sampling rate which is equal to 32768 and t2 =mod ((1:1:32768)/32768,1/20).

Next, EEMD is applied to the signal shown in **Fig. 2(c).** The resulting IMFs are shown in **Fig. 4**. **Fig. 4** indicates that impulses and low-frequency signals can be recovered from the resulting IMFs.

Further, CEEMD, CEEMDAN, and NPCEEMD are also applied to the signal shown in Fig. 2(c). The resulting IMFs are respectively shown in **Figs. 5-7**. From these Figs, we can observe that impulses and low-frequency signals can be recovered from the IMF produced by CEEMD and NPCEEMD. But in the IMFs produced by the CEEMDAN method, severe mode mixing can be seen in the low-frequency part.

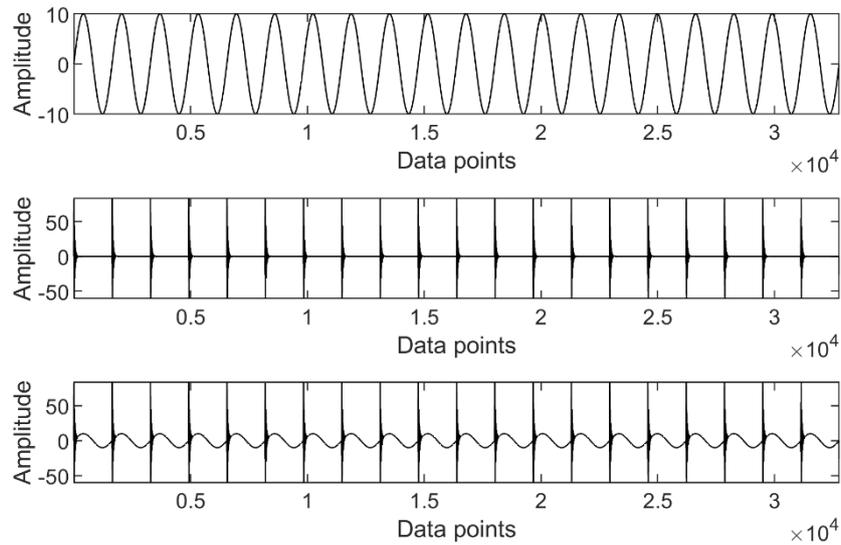

*Fig. 2. Simulating signal (a) Low-frequency response (b) Equal time-spaced impulses (c) Combined signal*

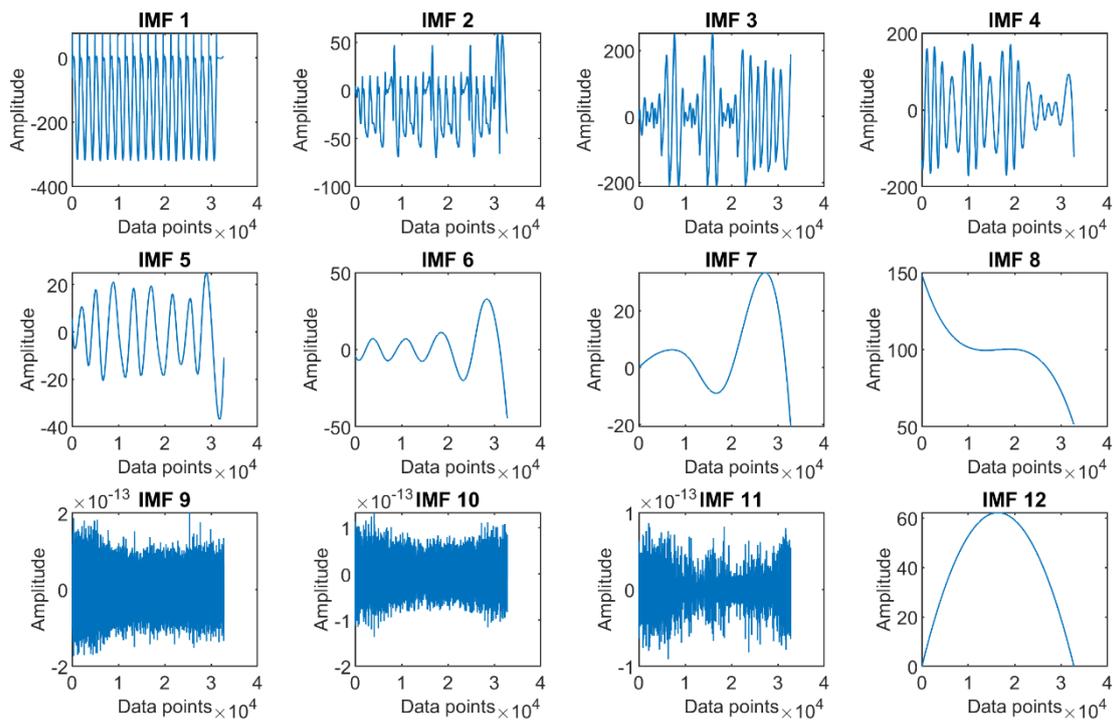

*Fig. 3. The IMFs obtained by the EMD method applied to the signal shown in Fig. 2(c)*

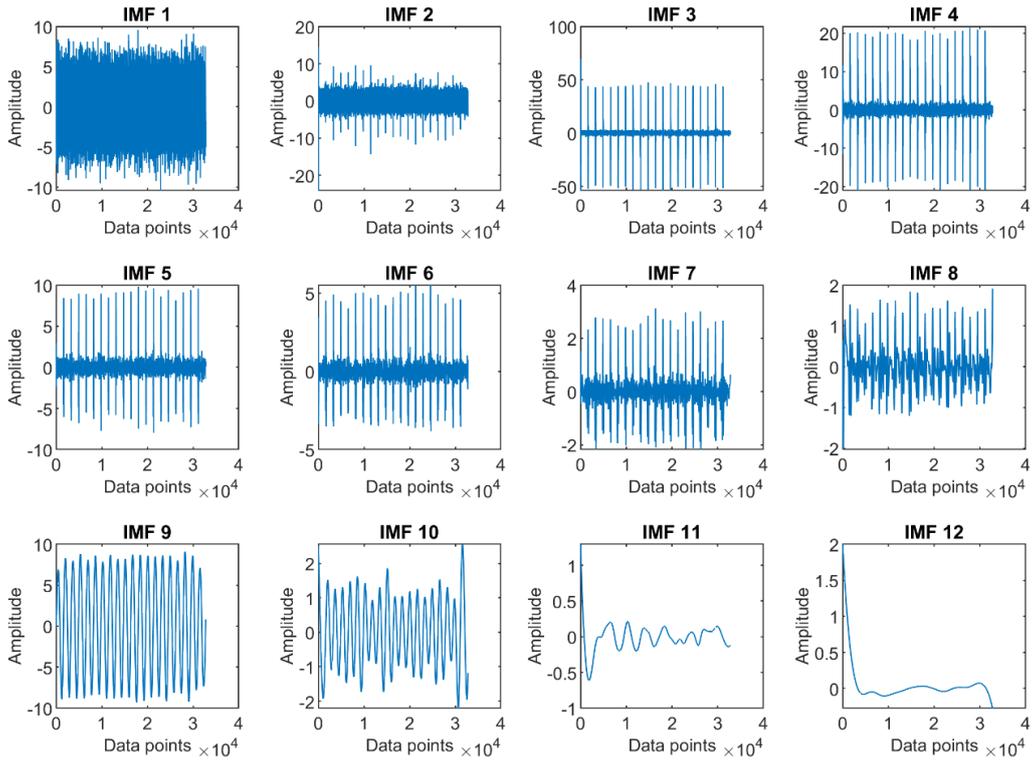

*Fig. 4. The IMFs obtained by the EEMD method applied to the signal shown in Fig. 2(c))*

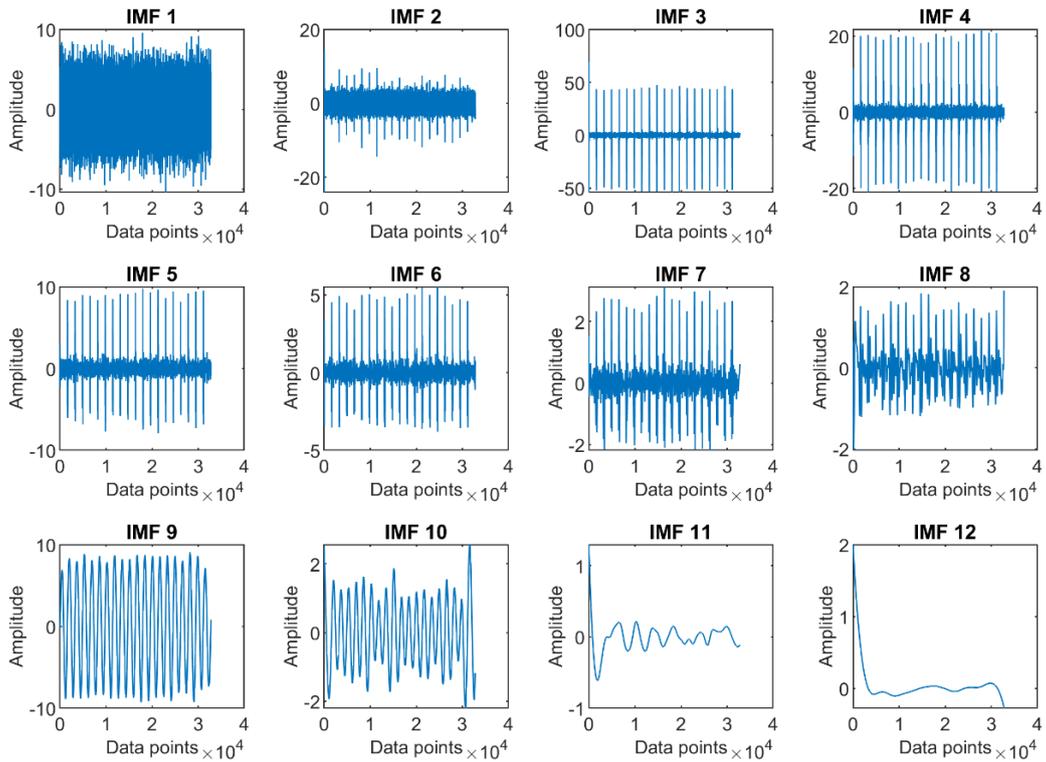

*Fig. 5. The IMFs obtained by the CEEMD method applied to the signal shown in Fig. 2(c)*

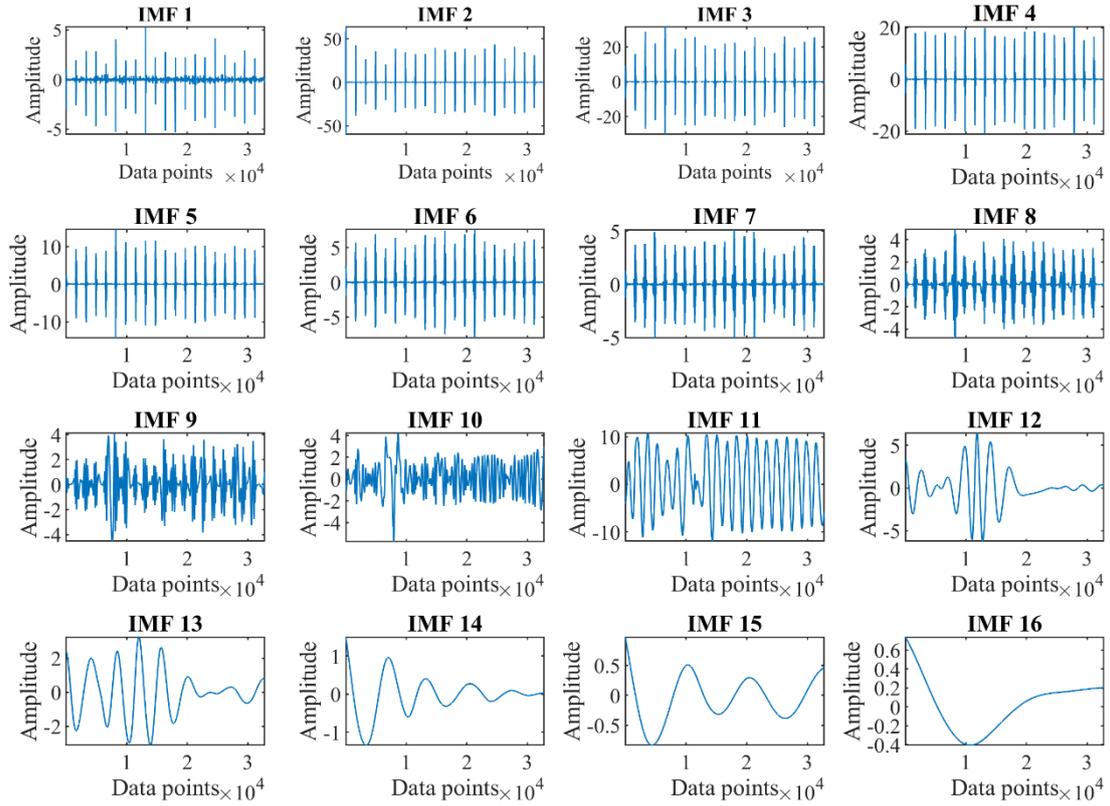

*Fig. 6. The IMFs obtained by the CEEMDAN method applied to the signal shown in Fig. 2(c)*

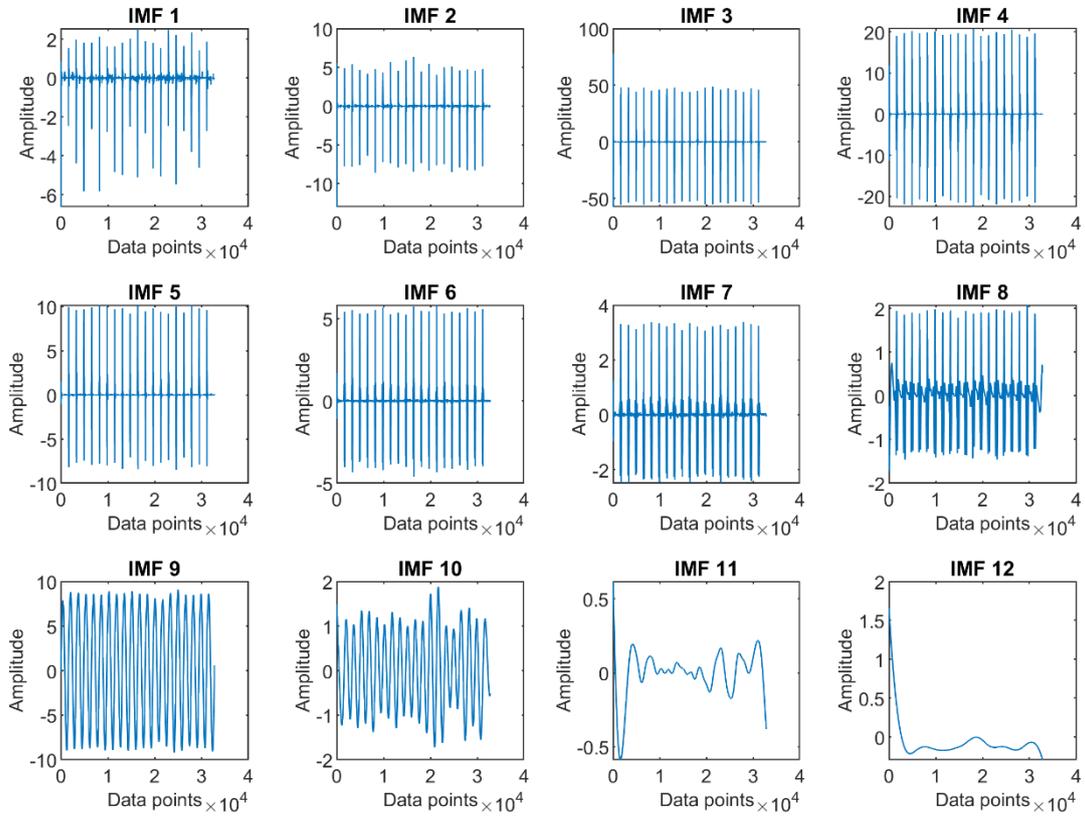

*Fig. 7. The IMFs obtained by the NPCEEMD method applied to the signal shown in Fig. 2(c)*

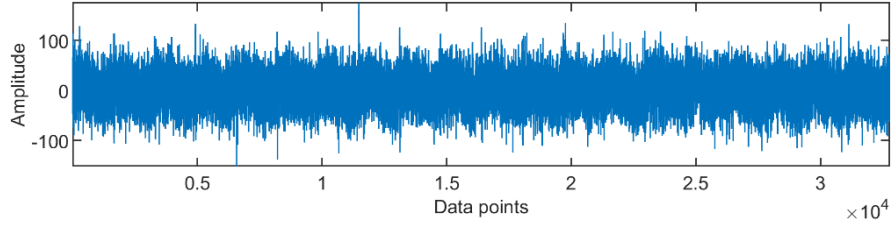

*Fig. 8 Signal of SNR equal to -30 dB (generated by adding noise to the signal in the Fig. 2(c))*

Next, noise is added to the signal shown in Fig. 2(c) in order to get the signal-to-noise (SNR) ratio equal to 30 dB. The resulting signal is shown in **Fig. 8**. The EMD, EEMD, CEEMD, CEEMDAN, and NPCEEMD methods are applied to the signal in the **Fig. 8**. The resulting IMFs are shown in **Figs. 9-13**. From these figures, it can be experienced that the performance of the existing EMD method is the worst in this case. In the case of CEEMD method, the pulses can be seen to spit out in the IMF3 and IMF4 due to the mode mixing effect. In the case of IMFs produced by the CEEMDAN method, severe mode mixing can be observed in the low-frequency part. However, using the proposed NPCEEMD method, the pulses, which can be found in the IMF4 and low-frequency signal, can also be easily recovered.

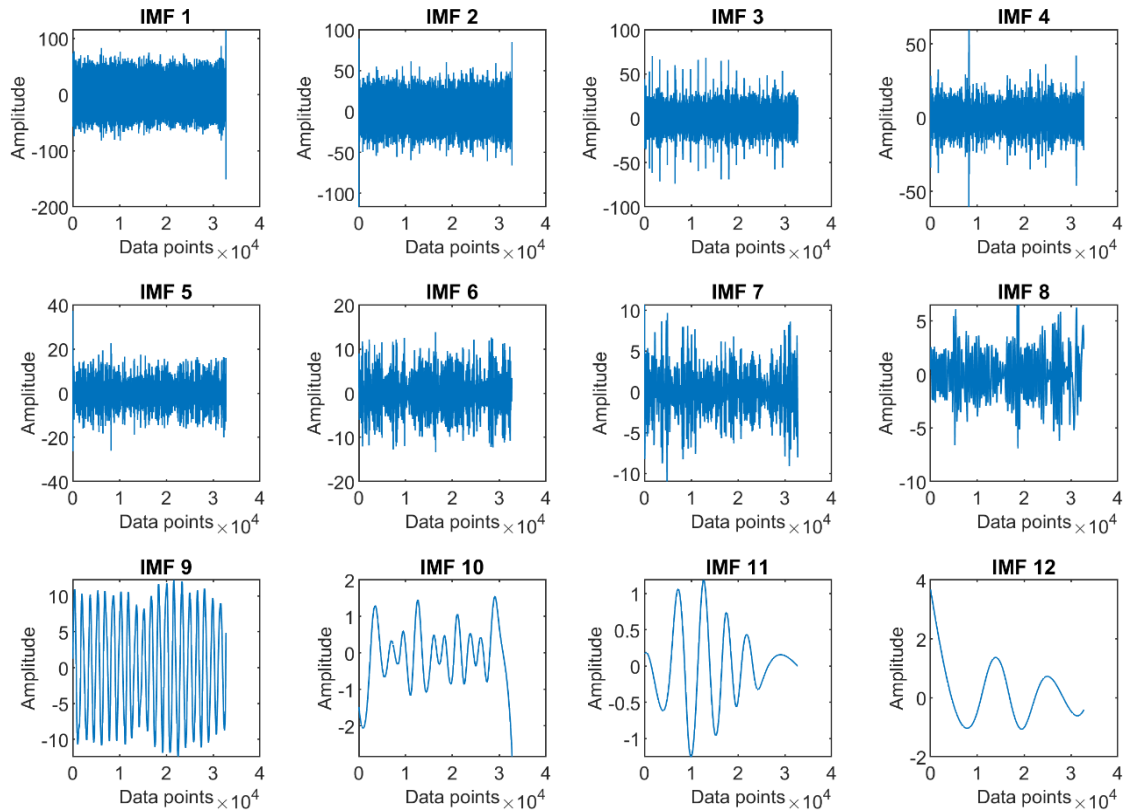

*Fig. 9 The IMFs obtained by the EMD method applied to the signal shown in Fig. 8*

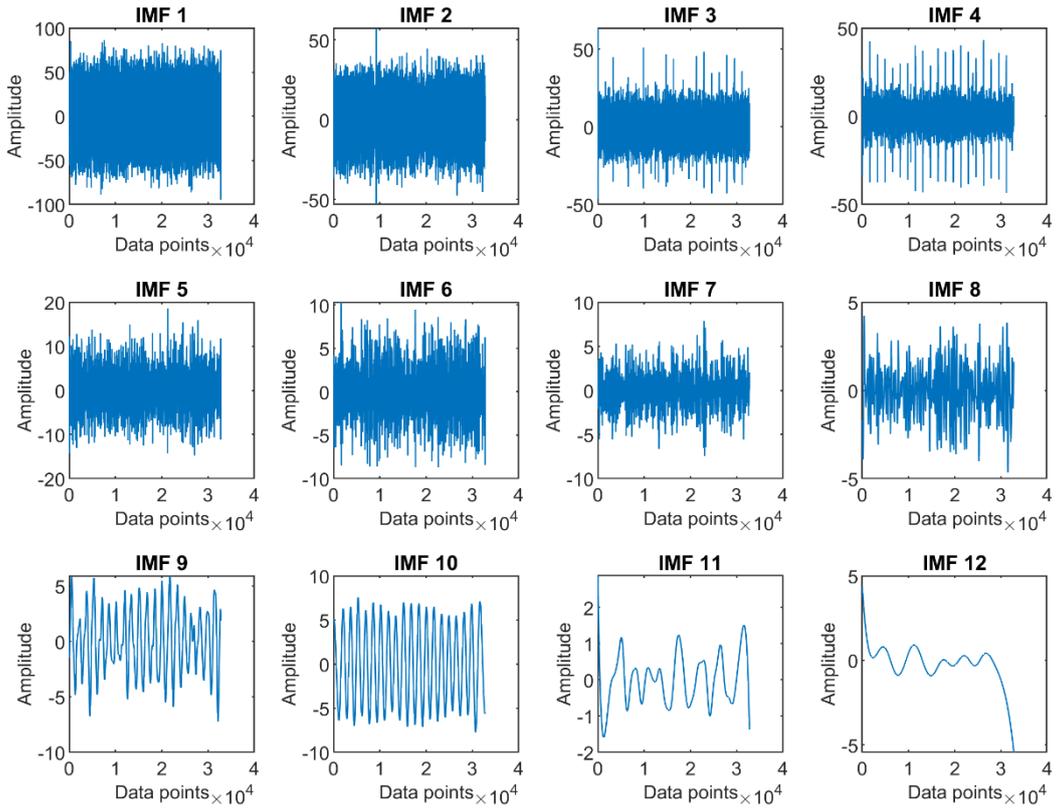

Fig. 10. *The IMFs obtained by the EEMD method applied to the signal shown in Fig. 8*

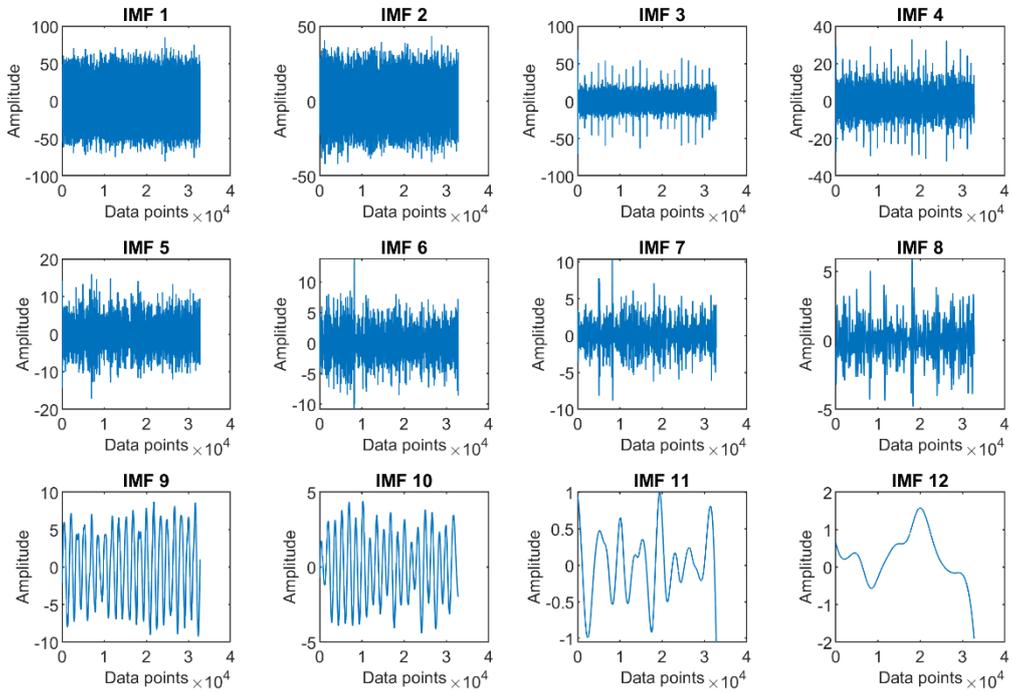

*Fig. 11. The IMFs obtained by the CEEMD method applied to the signal shown in Fig. 8*

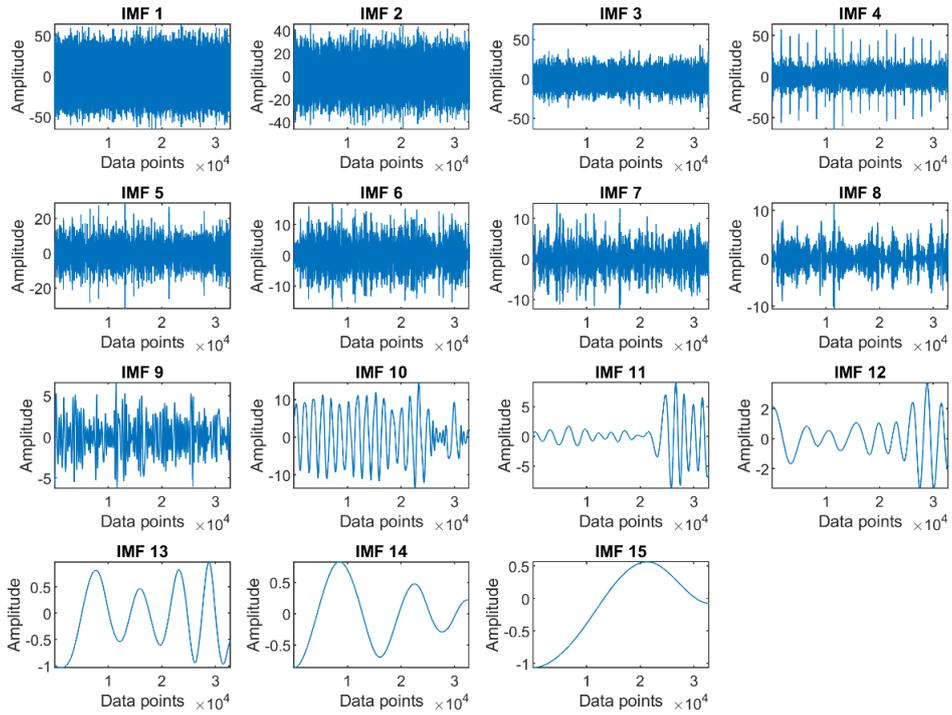

*Fig. 12. The IMFs obtained by the CEEMDAN method applied to the signal shown in Fig. 8*

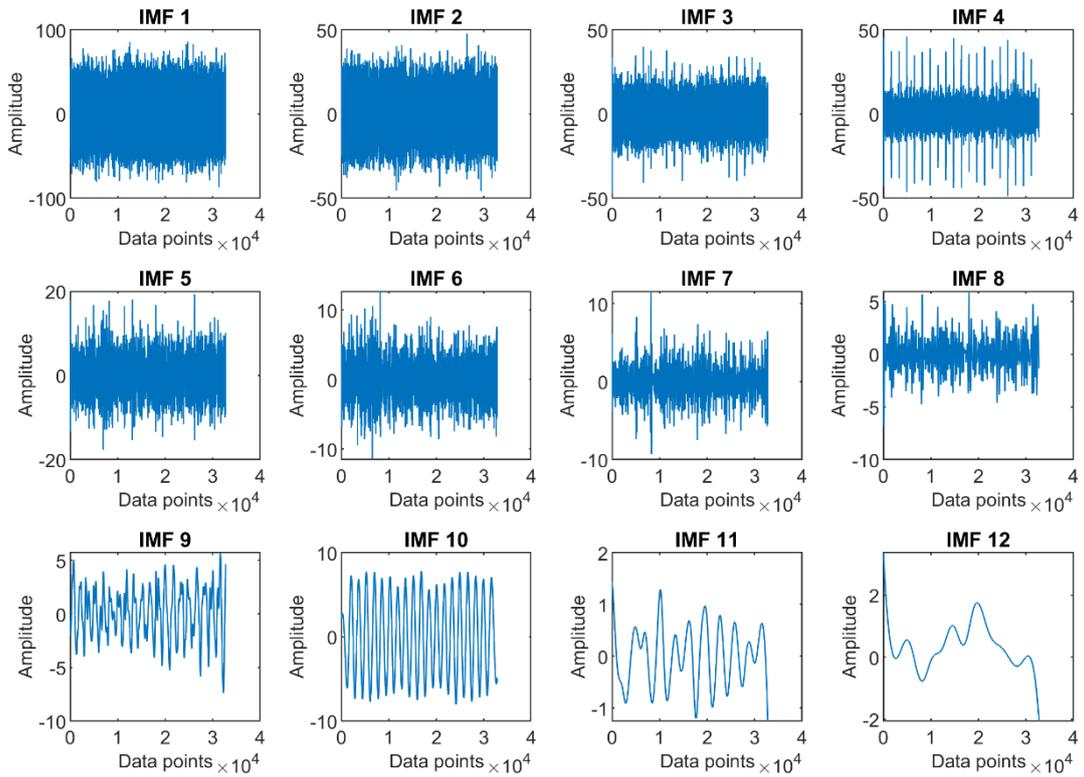

*Fig. 13. The IMFs obtained by the NPCEEMD method applied to the signal shown in Fig. 8*

The effect of the Hurst exponent on decomposition performance of the NPCEEMD method is examined in Appendix A.1. Our findings in the appendix A.1 show that as the H ratio increases from 0.1 to 0.9, mode mixing becomes start becoming prominent. Therefore, H=0.1 is suitable for decomposition. **In the Appendix A.2**, we have showed that the developed NPCEEMD performs well for extracting low and high frequency signal, at both ensemble sizes 10 and 100. The investigation indicates that any changes in NPCEEMD parameters have little effect on their performance. Therefore, analysis have been done at low ensemble numbers of 10 and H=0.1.The main purpose of this discussion is to justify the fact that with the changing conditions, parameters of the existing decomposition methods need to be tuned. On the other hand, the proposed NPCEEMD method is non-parametric and easily suits to every signal.

After evaluating the decomposition performance of the proposed NPCEEMD method, we next proceed to develop the proclaimed our non-parametric complementary ensemble empirical mode decomposition (NPCEEMD) based methodology to extract weak defect features for identifying bearing defects in rotary machinery.

## 4. Novel methodology for the identification of weak defect features

For reliable fault diagnosis, many existing vibration-based signal processing techniques are widely being used. However, the existing techniques have several limitations, including the mode mixing problem and high computational cost. To overcome these limitations, a nearly mode-mixing-free technique with minimal computational effort is developed as follows. **Fig. 14** displays the flowchart of the proposed NPCEEMD based methodology. The methodology consists of the following steps:

**Step 1:** Obtain the raw signal.
**Step 2:** Decompose the raw signal using the proposed NPCEEMD method.
**Step 3:** Use the nearest neighbor approach to compute the mutual information (MI) of the raw signal and the IMFs generated from the NPCEEMD method.
**Step 4:** A vibration signal consists of both noise and information. The noise is generally considered to be random and uncorrelated with the original waveform; as a result of which, the mutual information between the noise and original signal is less. The IMFs that have high mutual information with the raw signal is considered to be correlated and are supposed to have a high amount of mutual information. The IMFs with mutual information greater than 0.1 are combined to form the resulting IMF. The trial and error method is used to determine the value of mutual information. We have chosen a low value of 0.1 (10%) to capture the vast majority of the information (which means we intend to select the maximum number of IMFs). The value less than 0.1 indicates that the IMFs do not contain any information with useful content.

**Step 5:** The resulting IMFs are combined to construct a resulting signal.

**Step 6:** Finally, the envelope spectrum of the resulting IMFs signal is computed to confirm the presence of the defect.

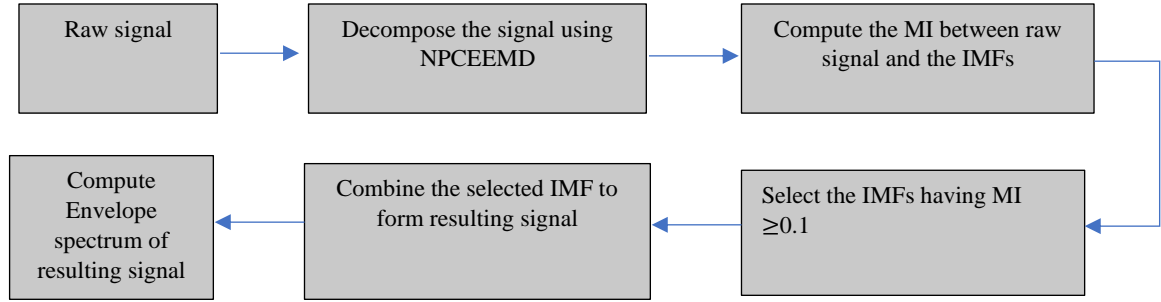

*Fig. 14. Flowchart of the proposed NPCEEMD method*

We shall now applicate the proposed non-parametric complementary ensemble empirical mode decomposition (NPCEEMD) based methodology of extracting the weak defect features for identifying bearing defects in rotary machinery.

## 5. Application of the proposed NPCEEMD based method

### 5.1 Simulation vibration response of a bearing in a run towards failure process

First, we apply the proposed methodology to the life-time failure data, generated using the simulation analysis.

The simulation vibration response of a bearing in a run towards failure process is obtained using equation (11)

$$x(t) = \sum_{i=1}^{i=M}\{A_i.\cos(2\pi f_m(i.T' + \xi_i))\}.\{e^{-B(t-iT-\xi_i)}\cos[2\pi f_n(t - iT' - \xi_i)]\} + n(t) \quad (11)$$

Here, $A_i$ is the amplitude modulator, $f_m$ is the shaft frequency, $f_n$ is the natural frequency, $T'$ is the period of impulses, $\xi_i$ is the minor and random fluctuation around the average period $T'$ and $B$ is an any factor to simulate the attenuation of oscillations waveform. The simulation vibration response is obtained by increasing the amplitude $A_i$ where $f_m$, $T'$, $f_n$, $A$, and sampling rate are 25 Hz, 0.015 s, 2000 Hz, 10 000 Hz, and 5, respectively, and $n(t)$ is white Gaussian noise. Each signal of duration 0.5 sec has been obtained after every 1 minute. A total of 500 specimens were generated, with a total lifetime of 500 minutes as shown in Fig. 15.

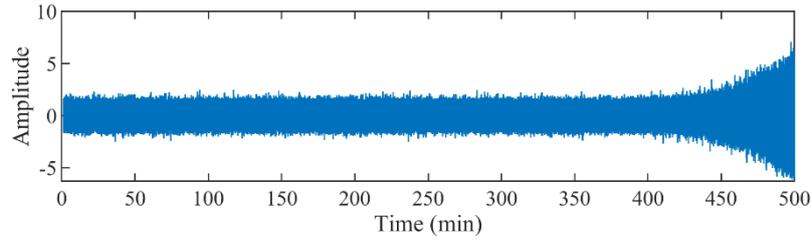

*Fig. 15 Simulation degradation data of 500 minutes*

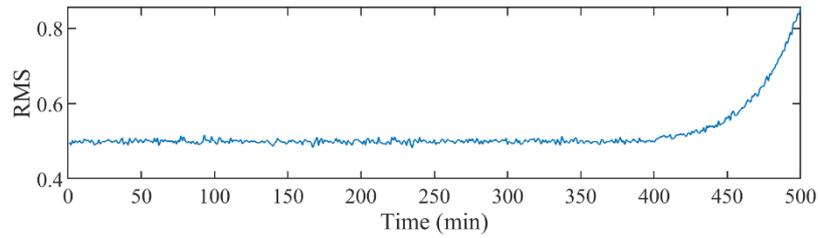

*Fig. 16. RMS of vibration signal presented in Fig. 15*

The RMS of data shown **in Fig. 15** is presented in **Fig. 16**. For fault identification, we choose the signal of 440$^{th}$ minute. The signal is shown in **Fig. 17**.

**5.1.1 Diagnosis Result of the proposed method**

**Fig. 18** shows the IMFs of the signal of 440$^{th}$ minute. The mutual information of the IMFs to the raw signal is shown in **Fig. 19**. IMF1, IMF2, and IMF3 have mutual information greater than 0.1. These IMFs are merged, and then the envelope spectrum of the resulting IMF is determined (**Fig. 20**). The defect frequency has the highest peak. Peaks at the harmonics of defect frequency can also be noticed in **Fig. 20**.

**Fig. 21** shows the envelope spectrum of the combined signal generated from the remaining IMF's. In this signal, no peak is observed at the defect frequency This indicates that the IMFs with mutual information less than 0.1 have no significance and can be ignored. Thus, the proposed criteria can be declared as a suitable tool for selecting the IMFs with defect features, or, in other words, a tool for determining the appropriate filtering bands.

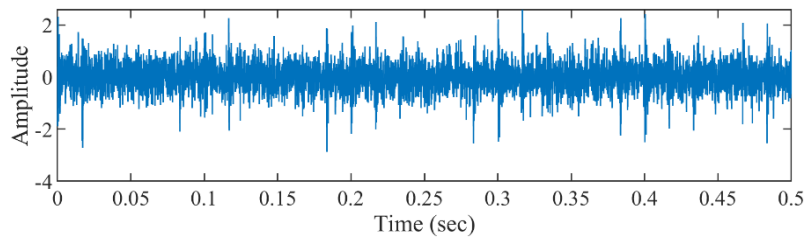

*Fig. 17. Vibration data of 440 minutes*

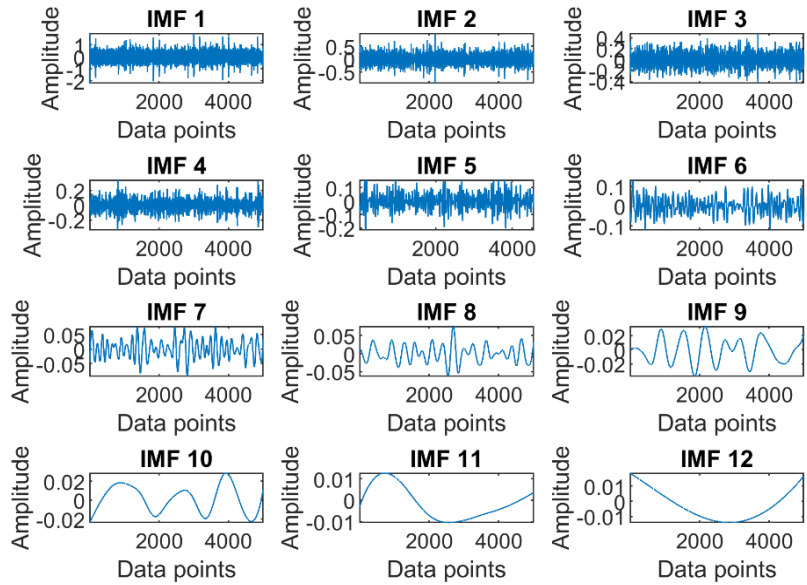

*Fig. 18. IMFs generated by applying NPCEEMD to the signal shown in Fig. 17*

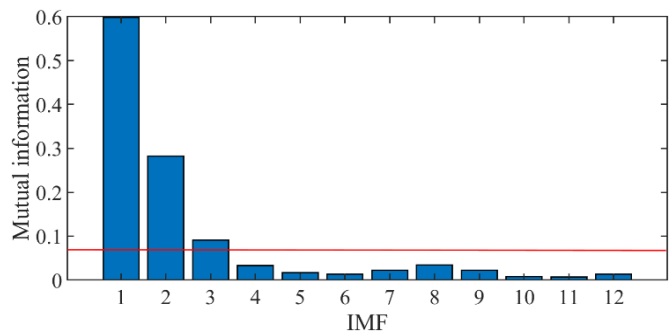

*Fig. 19. Mutual information of IMFs (Fig. 18) to the raw signal (Fig. 17) of File number 534*

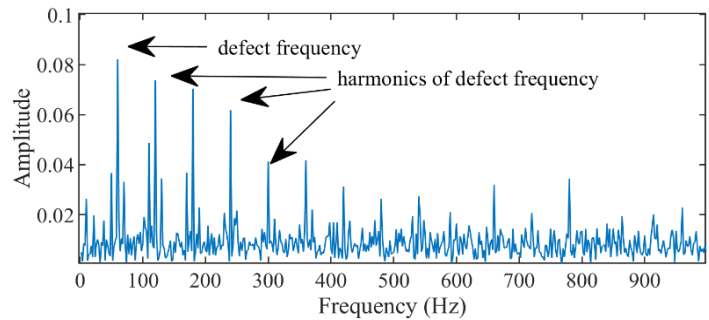

*Fig. 20. Envelope spectrum of the combined signal generated using IMF1, IMF2, and IMF3*

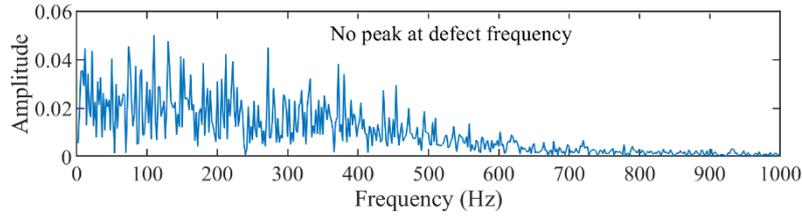

*Fig. 21. Envelope spectrum of the combined signal generated using remaining IMFs*

**5.1.2 Diagnosis Result of the EEMD-based method**

A comparison of the proposed NPCEEMD based methodology and the existing EEMD method was also carried out. **Fig. 22** displays the IMFs generated by applying the EEMD to the signal shown in **Fig. 17**. Kurtosis is a statistical parameter that indicates signal impulsiveness and is commonly used to select the band containing defect-related information. For the selection of IMFs having defect-related information, the kurtosis value of various IMFs is computed. The kurtosis of IMFs shown in **Fig. 22** is displayed in **Fig. 23**. The IMF 11 possesses the greatest kurtosis value. Consequently, the envelope spectrum of IMF 11 is computed and displayed in **Fig. 24**. No Peak is found at the bearing defect frequency. The performance of the existing EEMD, CEEMD, CEEMDAN and the proposed NPCEEMD methods under different cases has been mentioned in **Table 1**.

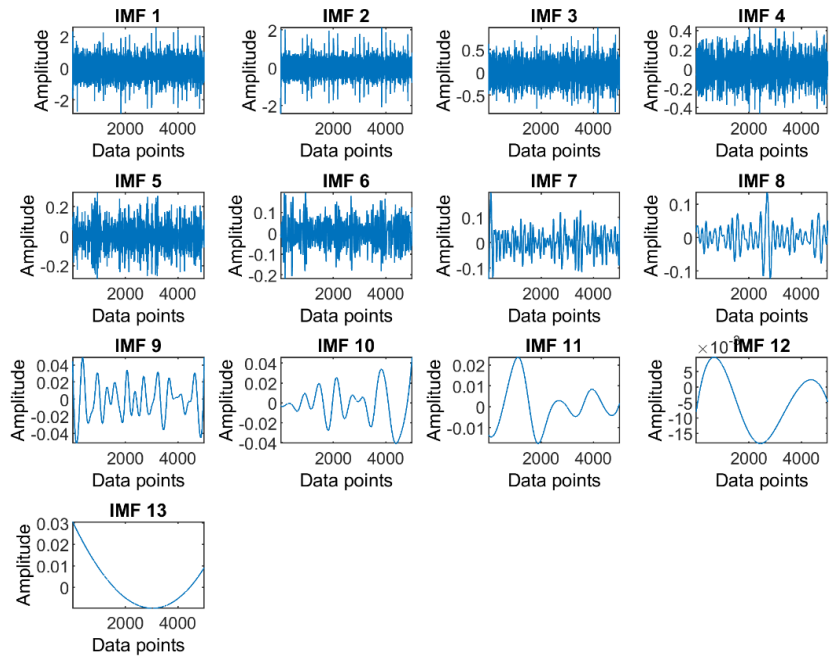

*Fig. 22. IMFs generated by applying EEMD to the signal shown in Fig. 17*

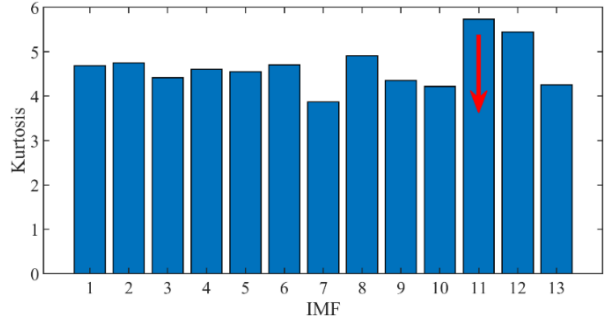

*Fig. 23. Kurtosis of IMFs shown in Fig. 22*

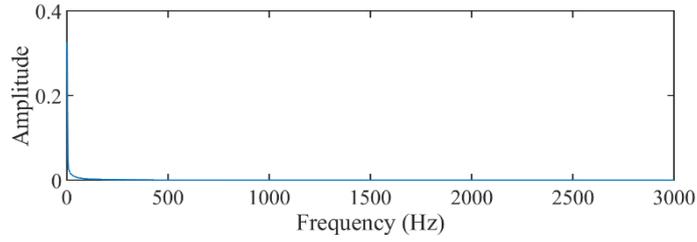

*Fig. 24. Envelope spectrum of IMF11 (showed in Fig. 22)*

Table 1: Performance of the EEMD, CEEMD, CEEMDAN and the proposed NPCEEMD methods

| Case | Performance of existing method | | | Performance of the proposed method |
|---|---|---|---|---|
| | EEMD | CEEMD | CEEMDAN | |
| 5.1 Simulation vibration response of a bearing in a run towards failure process | Defect could not be identified | Defect could be identified | Defect could not be identified | Defect could be identified |
| 5.2 Early identification of bearing defect during degradation (data provided by Intelligent Maintenance Systems (IMS) centre) | Defect could not be identified | Defect could be identified | Defect could not be identified | Defect could be identified |
| 5.3 Early identification of bearing defect during degradation (data provided by XJTU) | Defect could not be identified | Defect could be identified | Defect could be identified | Defect could be identified |
| 5.4 Bearing defect identification in the axial piston pump | Defect could be identified | Defect could be identified | Defect could be identified | Defect could be identified |
| 5.5 Bearing defect identification using acoustic data | Defect could be identified | Defect could be identified | Defect could be identified | Defect could be identified |

## 5.1.3 Diagnosis Result of the CEEMD-based method

**Fig. 25** displays the IMFs generated by applying the CEEMD to the signal shown in **Fig. 17**. For the selection of IMFs having defect-related information, the kurtosis value of various IMFs is computed. The kurtosis value of IMFs shown in **Fig. 25** is displayed in **Fig. 26**. The IMF 1 possesses the greatest kurtosis value. Consequently, the envelope spectrum of IMF 1 is computed and displayed in **Fig. 27**. A peak is found at the bearing defect frequency. This means that the performance of the CEEMD-based method in this instance can be stated as good.

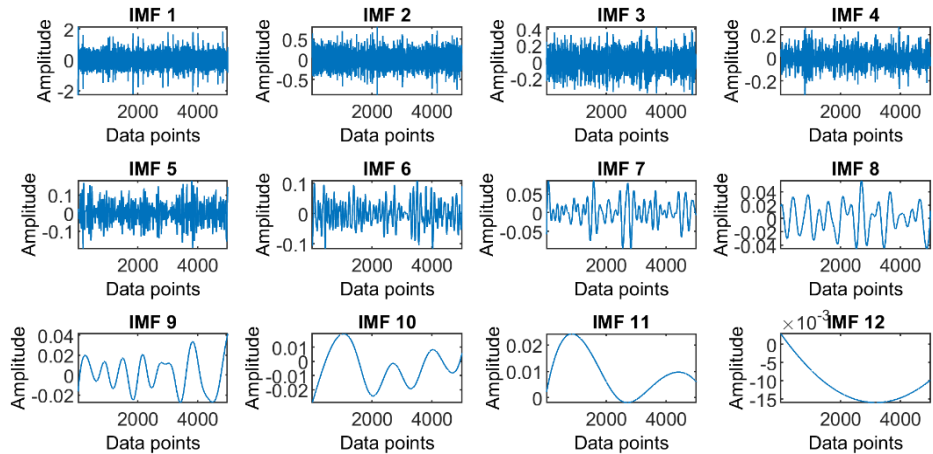

*Fig. 25. IMFs generated by applying CEEMD to the signal shown in Fig. 17*

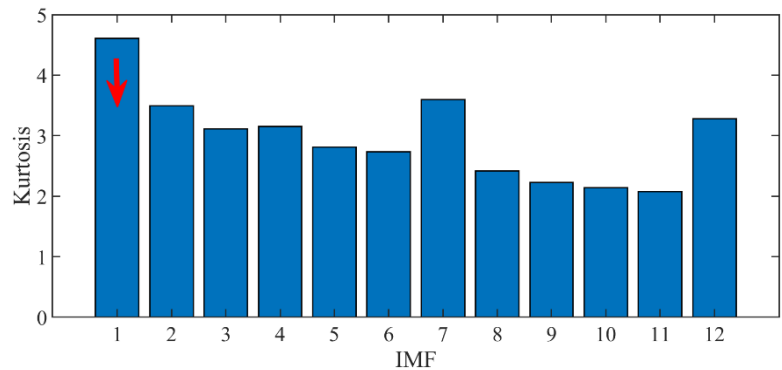

*Fig. 26. Kurtosis of IMFs shown in Fig. 25*

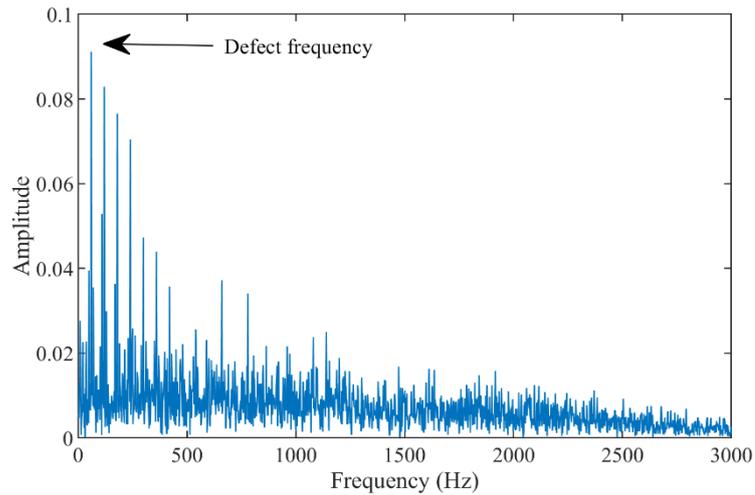

*Fig. 27. Envelope spectrum of IMF1 (showed in Fig. 25)*

### 5.1.4 Diagnosis Result of the CEEMDAN-based method

**Fig. 28** shows the IMFs generated by applying the CEEMDAN method to the signal shown in **Fig. 17**. For the selection of IMFs having defect-related information, the kurtosis value of various IMFs is computed. The kurtosis values of the IMFs shown in **Fig. 28** are displayed in **Fig. 29**. The IMF 7 possesses the greatest kurtosis value. Consequently, the envelope spectrum of IMF 7 is computed and displayed in **Fig. 30**. No peak is found at the bearing defect frequency. This indicates that the performance of the existing CEEMDAN-based method in this instance is not good.

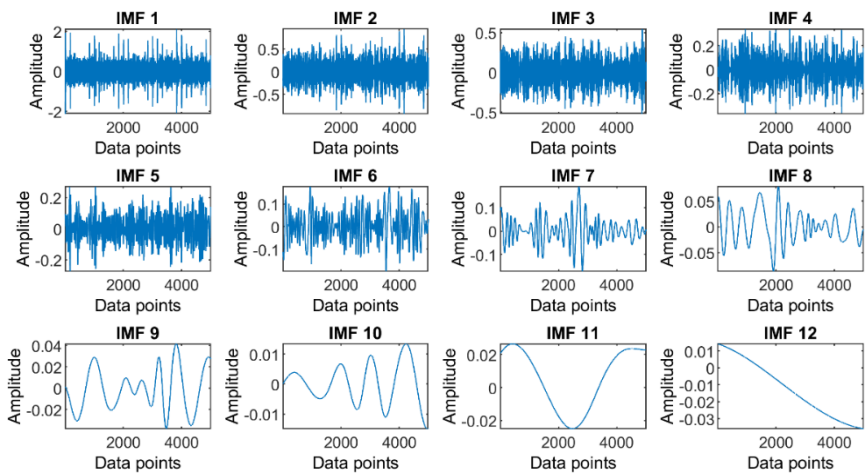

*Fig. 28. IMFs generated by applying CEEMDAN to the signal shown in Fig. 17*

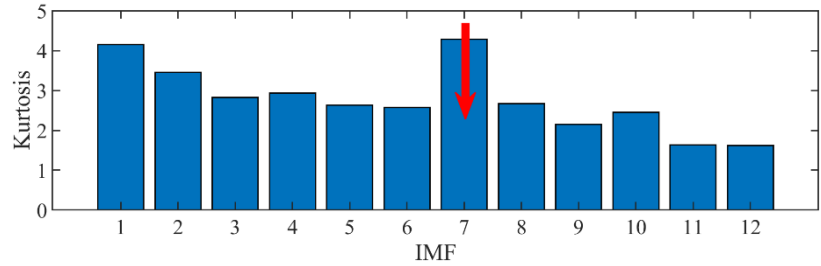

*Fig. 29. Kurtosis of IMFs shown in Fig. 28*

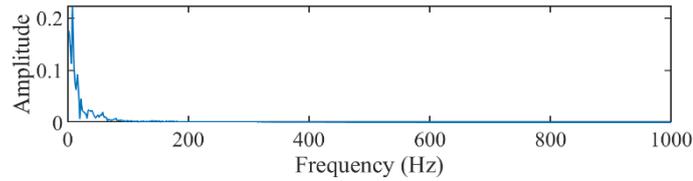

*Fig. 30. Envelope spectrum of IMF 7 (showed in Fig. 28)*

## 5.2 Early identification of bearing defect during degradation (data provided by Intelligent Maintenance Systems (IMS) centre)

In this section, the proposed NPCEEMD method is applied to evaluate the acceleration data set, provided by the National Science Foundation I/UCR Centre for Intelligent Maintenance Systems (IMS) [37,38]. The ICP accelerometers were installed on the bearing housing. **Fig. 31** depicts a schematic depiction of the test rig. The shaft is held in place by four bearings. An alternating current motor powered the bearing shaft via a pulley and belt system. The testing was done under constant load and speed conditions.

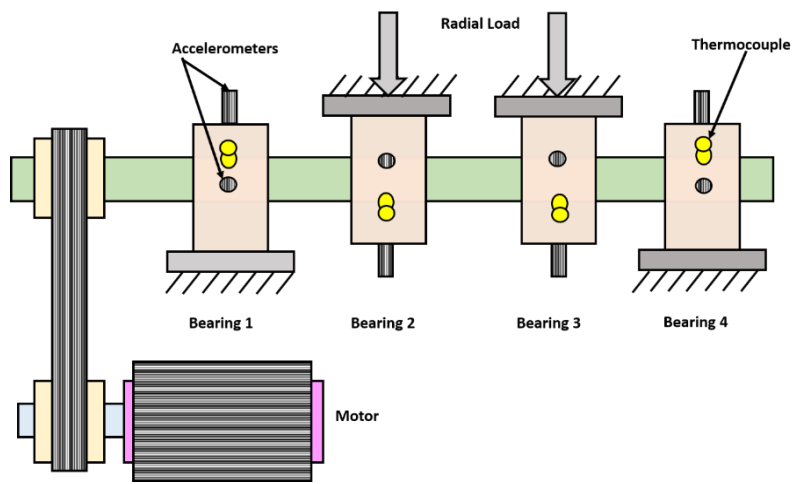

*Fig. 31. Bearing accelerated life test rig of the centre for intelligent maintenance systems*

The data in **Fig. 32** consist of lifetime failure data for a bearing. The vibration data was acquired at a sampling frequency of 20,000 samples per second. Each measurement consists of 20480 samples, which is equivalent to 1.02 seconds. The time interval between the two measurements was 10 minutes. **Fig. 33** shows the RMS of the signal shown in **Fig. 32**. At 5340 minutes (file no. 534), the RMS shows an upward trend. For a detailed investigation, the proposed signal processing scheme is applied to the vibration data. The vibration signal of file number 534 is shown in **Fig. 34**.

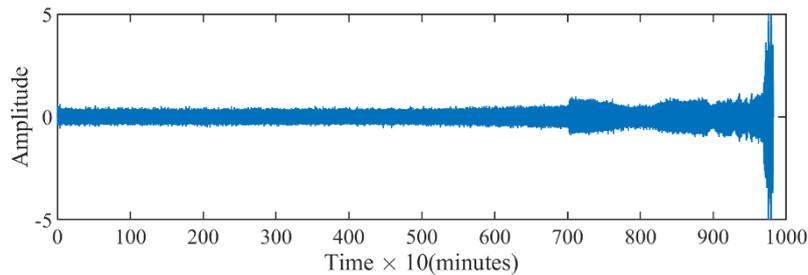

Fig. 32. Life time failure data of bearing (IMS data set)

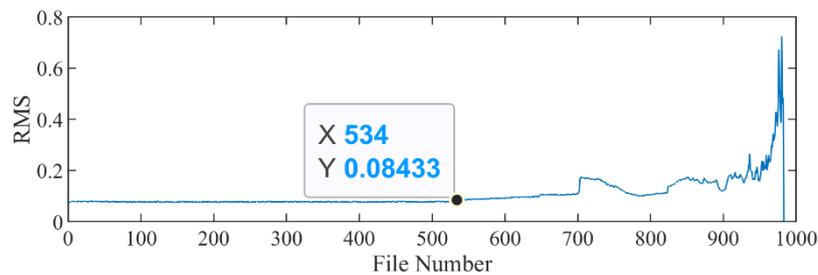

Fig. 33. RMS of vibration signal presented in Fig. 32

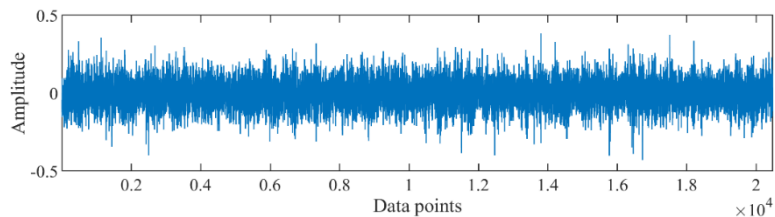

Fig. 34. Vibration data of file number 534

### 5.2.1 Diagnosis Result of the NPCEEMD-based method

**Fig. 35** shows the IMFs of the signal of file number 534 (**Fig. 34**). The mutual information of the IMFs to the raw signal is shown in Fig. 36. The IMF1, IMF2, and IMF3 has mutual information value greater than 0.1. These IMFs are merged and thereafter the envelope spectrum of the combined signal generated using IMF1, IMF2, and IMF3 determined (shown in **Fig. 37**). The bearing pass frequency outer race (BPFO) has the highest peak. Peaks at the BPFO can also be noticed in **Fig. 37**.

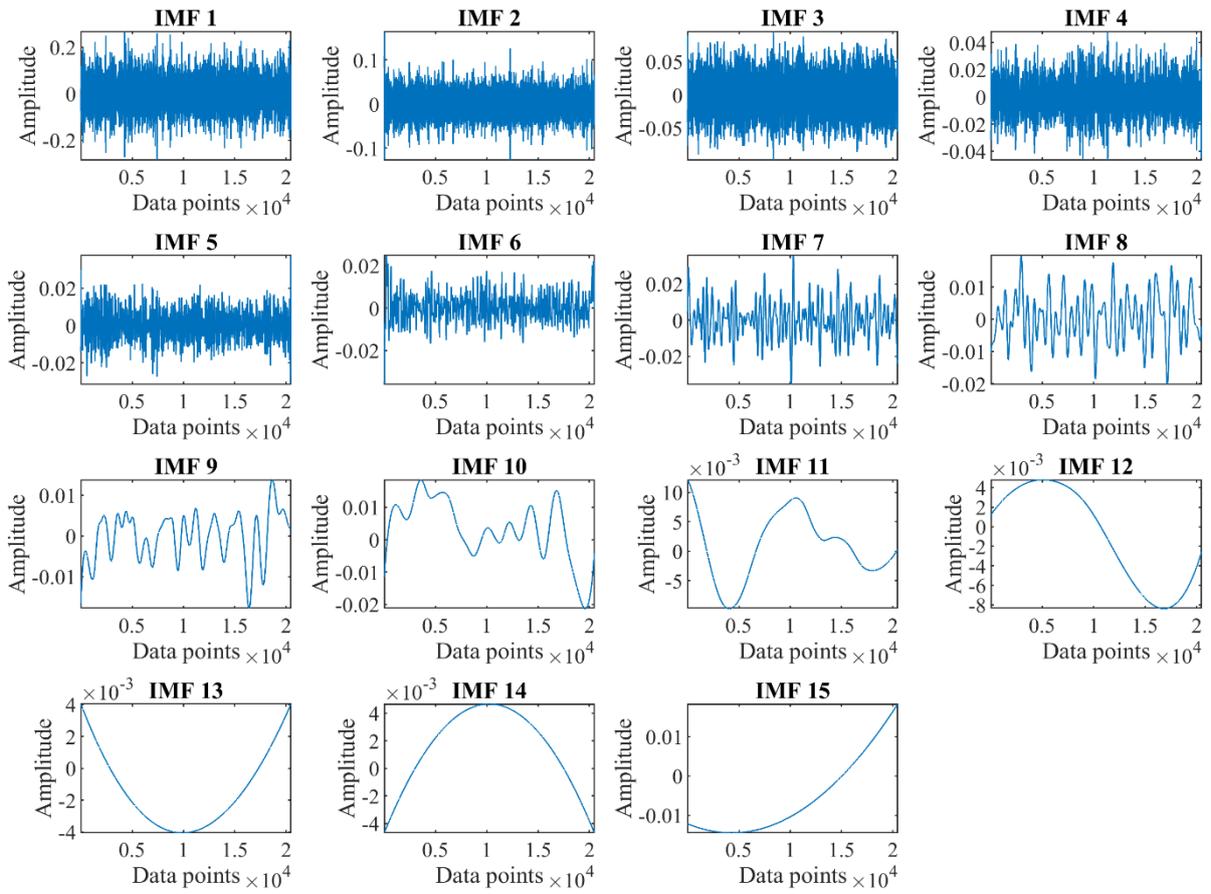

*Fig. 35. IMFs of the signal of file number 534 computed using NPCEEMD*

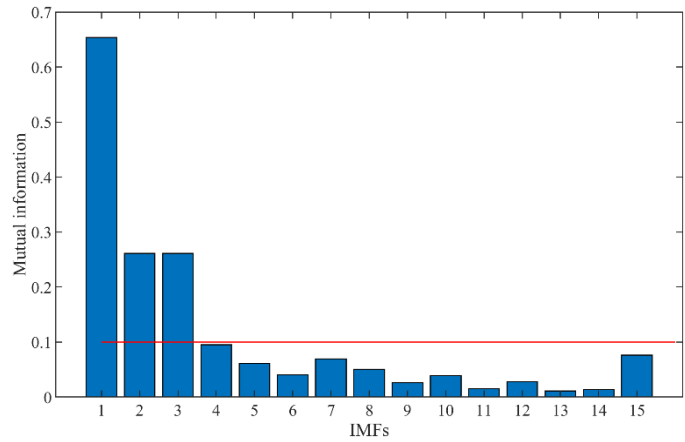

*Fig. 36. Mutual information of IMFs (Fig. 35) to the raw signal (Fig. 34) of File number 534*

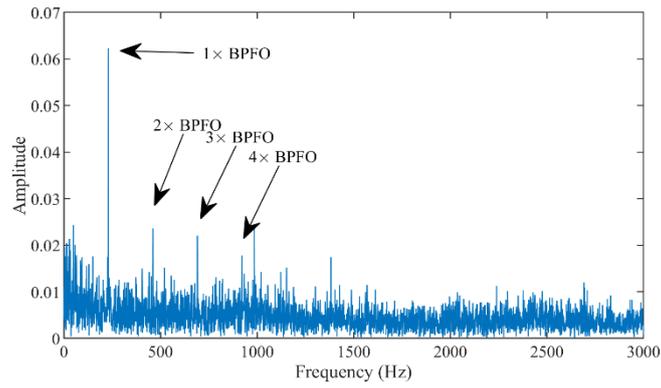

*Fig. 37. Envelope spectrum of the combined signal generated using IMF1, IMF2, and IMF3*

## 5.2.2 Diagnosis Result of the EEMD-based method

**Fig. 38** shows the IMFs generated by applying the EEMD method to the signal shown in **Fig. 34.** For the selection of IMFs having defect-related information, the kurtosis value of various IMFs is computed. The kurtosis of IMFs shown in **Fig. 38** is displayed in **Fig. 39**.

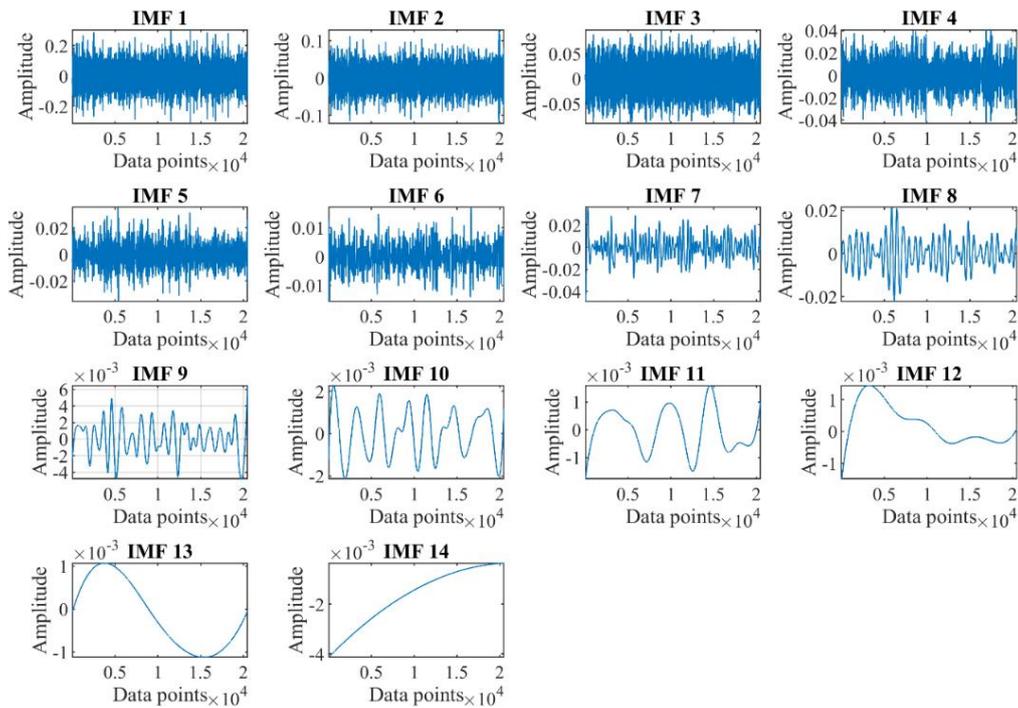

*Fig. 38. IMFs generated by applying the EEMD to the signal shown in Fig. 34*

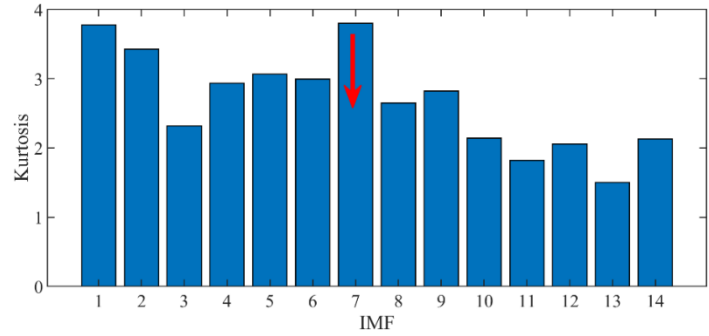

*Fig. 39. Kurtosis of IMFs shown in the Fig. 38*

The IMF 7 possesses the greatest kurtosis value. Consequently, the envelope spectrum of IMF 7 is computed and displayed in **Fig.** . No peak is found at the bearing defect frequency. This means that the performance of the EEMD-based method in this instance is poor.

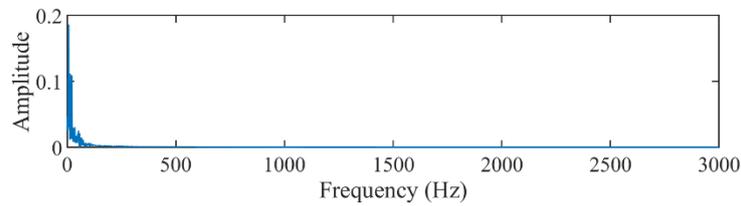

*Fig. 40. Envelope spectrum of IMF7(showed in Fig. 38)*

### 5.2.3 Diagnosis Result of the CEEMD-based method

**Fig. 41** shows the IMFs generated by applying the CEEMD method to the signal shown in **Fig. 34**. The kurtosis of IMFs shown in Fig. 41 is displayed in **Fig. 42**. The IMF 1 possesses the greatest kurtosis value. Consequently, the envelope spectrum of IMF 1 is computed and displayed in **Fig. 43**. Peak is found at the bearing defect frequency. This shows that the performance of the CEEMD-based method in this instance is good.

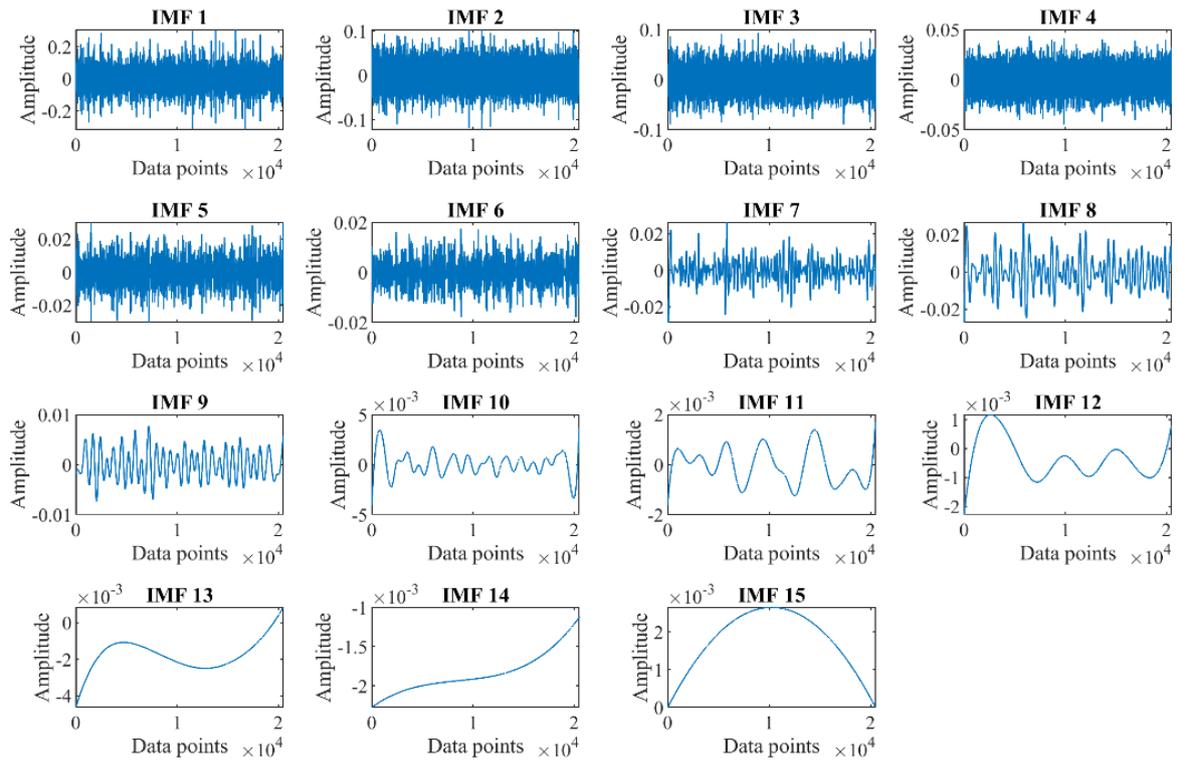

*Fig. 41. IMFs generated by applying the CEEMD method to the signal shown in Fig. 34*

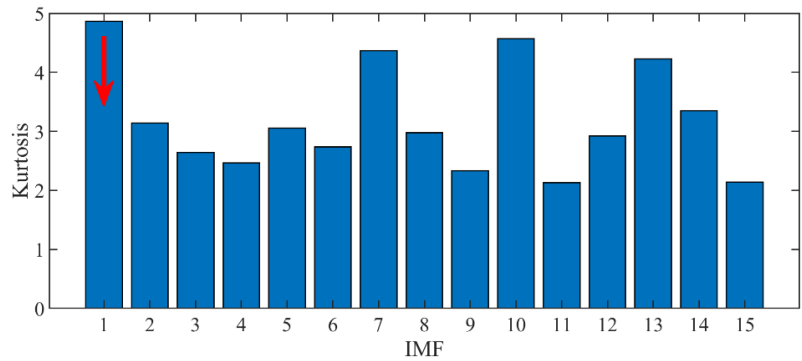

*Fig. 42. Kurtosis of the IMFs shown in Fig. 41*

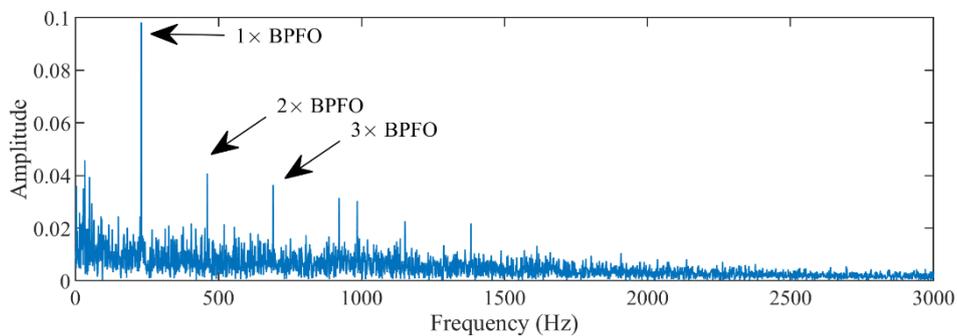

*Fig. 43. Envelope spectrum of the IMF1(showed in Fig. 41)*

## 5.2.4 Diagnosis Result of the CEEMDAN-based method

**Fig. 44** shows the IMFs generated by applying the CEEMDAN method to the signal shown in **Fig. 34.** The kurtosis of IMFs shown in Fig. 44 is displayed in **Fig. 45**. The IMF 8 possesses the greatest kurtosis value. Consequently, the envelope spectrum of IMF 8 is computed and displayed in **Fig. 46**. Peak couldn't be found at the bearing defect frequency. This shows that the performance of the CEEMDAN-based method in this instance is not good.

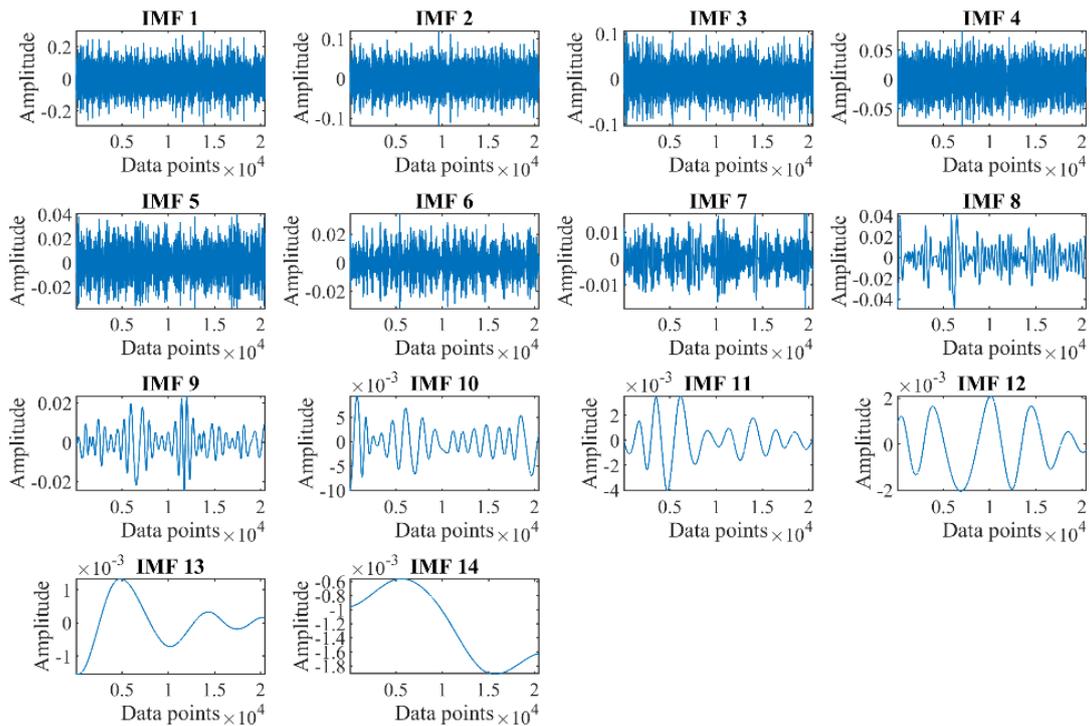

*Fig. 44. IMFs generated by applying the CEEMDAN method to the signal shown in Fig. 34*

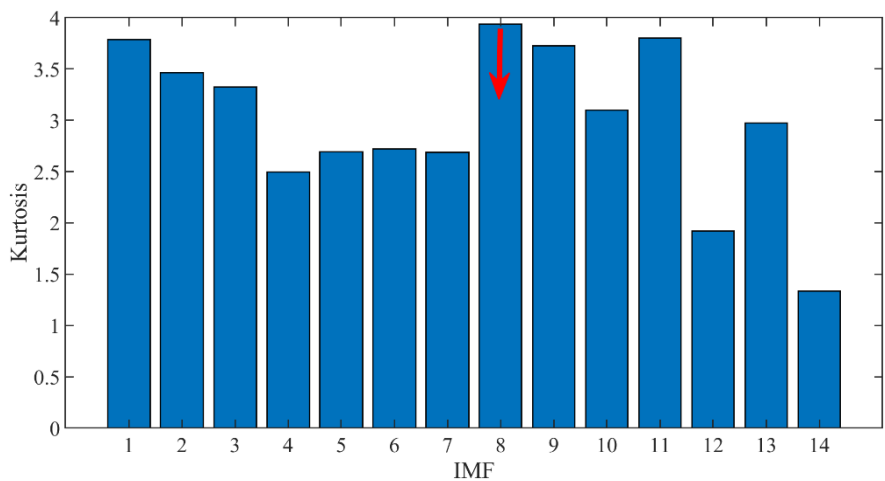

*Fig. 45. Kurtosis of the IMFs shown in Fig. 44*

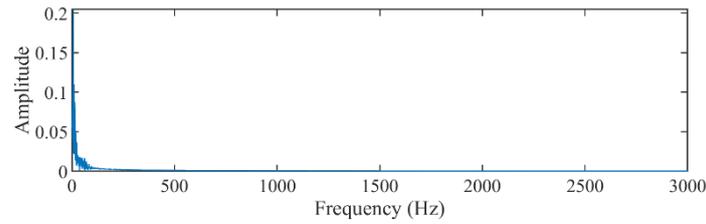

*Fig. 46. Envelope spectrum of the IMF8 (showed in Fig. 44)*

## 5.3 Early identification of bearing defect during degradation (data provided by XJTU)

The proposed NPCEEMD based method is employed to evaluate the accelerated data set, publicly made available by Xi'an-Jiaotong University (XJTU), China [39].

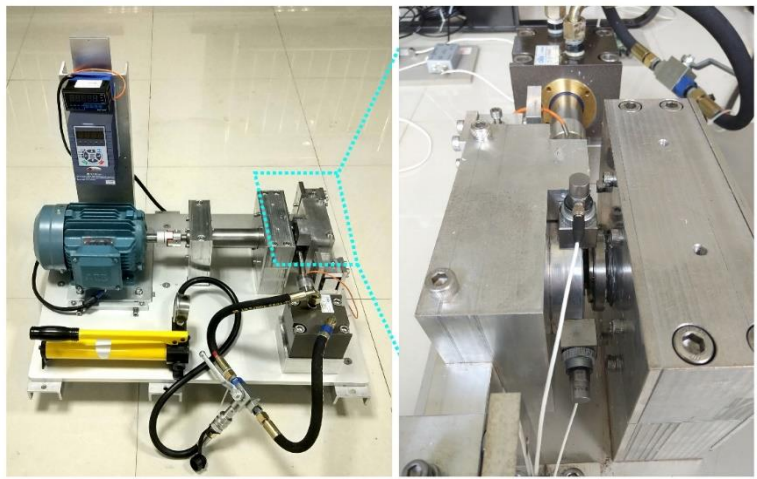

*Fig. 47. Bearing Accelerated life test rig of XJTU-SY*

**Fig.** depicts a bearing test bed that is equipped with a rolling element bearing (LDK UER204), an AC induction motor, a VFD speed controller, a shaft, a hydraulic loading arrangement, and a VFD speed controller. This is to guarantee that the rolling element bearing degrades quickly. Therefore, to comply with the aforementioned standard, the rolling element was put through a series of tests in which it was subjected to varying loads and speeds on the shaft. Two ICP accelerometers were mounted at 90-degree angles on the bearing housing and used to record vibration signals; one vibration sensor was measured in the horizontal plane, and the other was measured in the vertical plane. The sampling frequency was set at 25.6 kHz. The sampling period between each pair of measurements was one minute. Each measurement consists of 32768 samples, which is equivalent to 1.28 seconds.

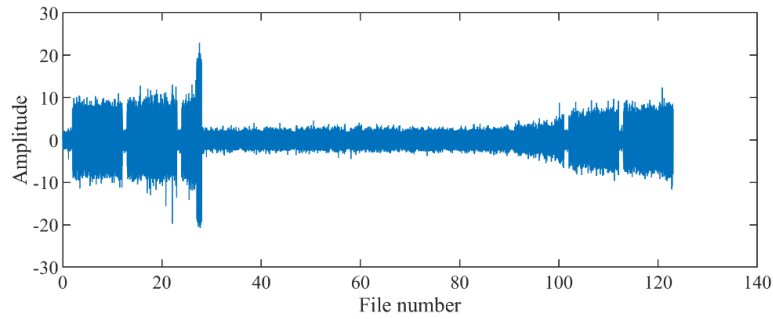

*Fig. 48. Life time failure data of bearing (XJTU data set)*

The data in **Fig. 48** are lifetime failure data for bearings from the XJTU data set. The RMS of the signal shown in Fig. 48 is presented in **Fig. 49**. The RMS shows an upward trend around 100 minutes. However, defects occur well before the 100th minute. The signal of the 79th minute is investigated at random. The vibration signal of file number 79 is shown in **Fig. 50**. For a detailed investigation, the proposed signal processing scheme is applied to the vibration data.

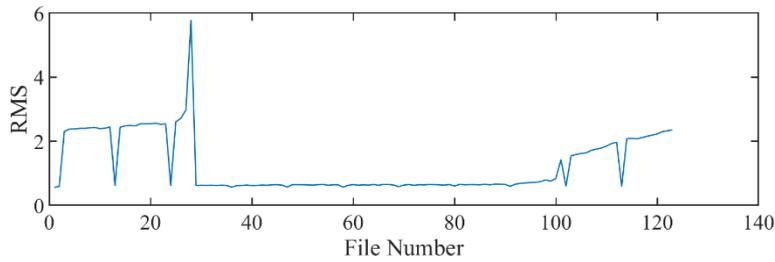

*Fig. 49. RMS of vibration signal presented in Fig. 48*

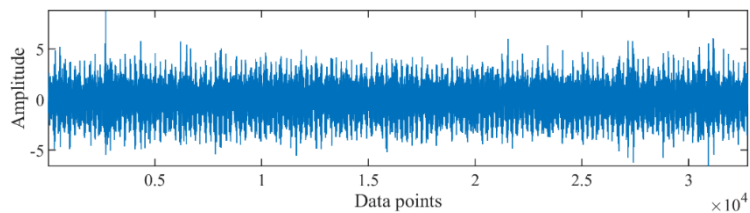

*Fig. 50. Vibration data of File number 79*

### 5.3.1 Diagnosis Results of the proposed NPCEEMD-based method

**Fig. 51** shows the IMFs of the signal in file number 79. The mutual information of the IMFs to the raw signal is shown in **Fig. 52**. The IMF1, IMF3, and IMF4 have mutual information value greater than 0.1. These IMFs are combined, and their envelope spectra are computed and shown in **Fig. 53**. A high peak can be seen at the bearing pass frequency outer race (BPFO). Peaks at the harmonic of BPFO can also be seen in **Fig. 54**.

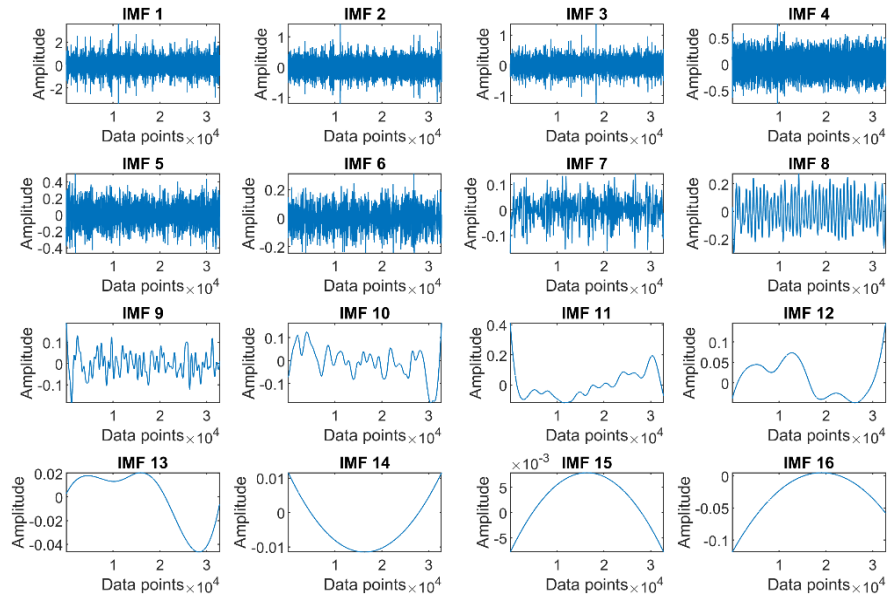

*Fig. 51. IMFs generated by applying the NPCEEMD method to the signal of file number 79*

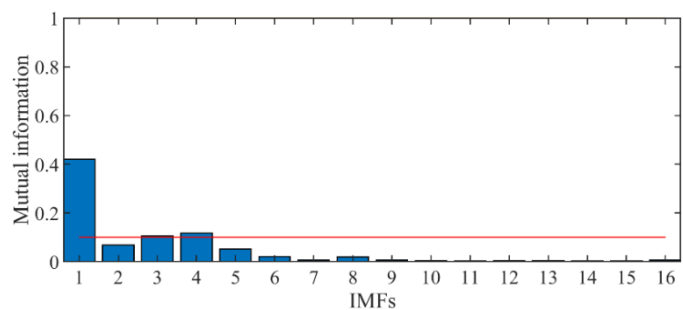

*Fig. 52. Mutual information of IMFs (Fig. 51) to the raw signal of file number 79 (Fig. 50)*

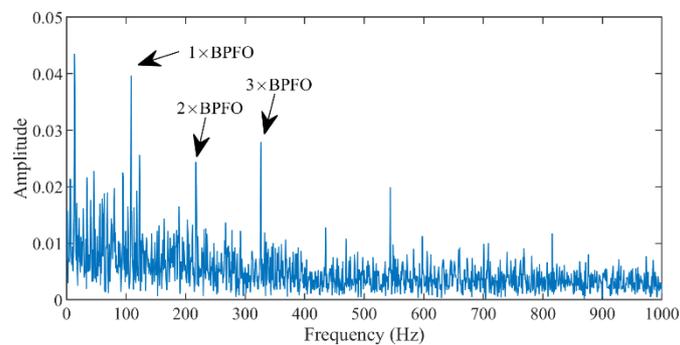

*Fig. 53. Envelope spectrum of the combined signal generated using IMF1, IMF3, and IMF4*

## 5.3.2 Diagnosis Results of the EEMD-based method

**Fig. 54** shows the IMF generated by applying the EEMD method to the signal shown in **Fig. 50**. The kurtosis of IMFs shown in **Fig. 54** is given in **Fig. 55**. The IMF 10 has the highest kurtosis value. Therefore, the envelope spectrum of IMF 10 is computed and shown in **Fig. 56.** No peak is found at the bearing defect frequency.

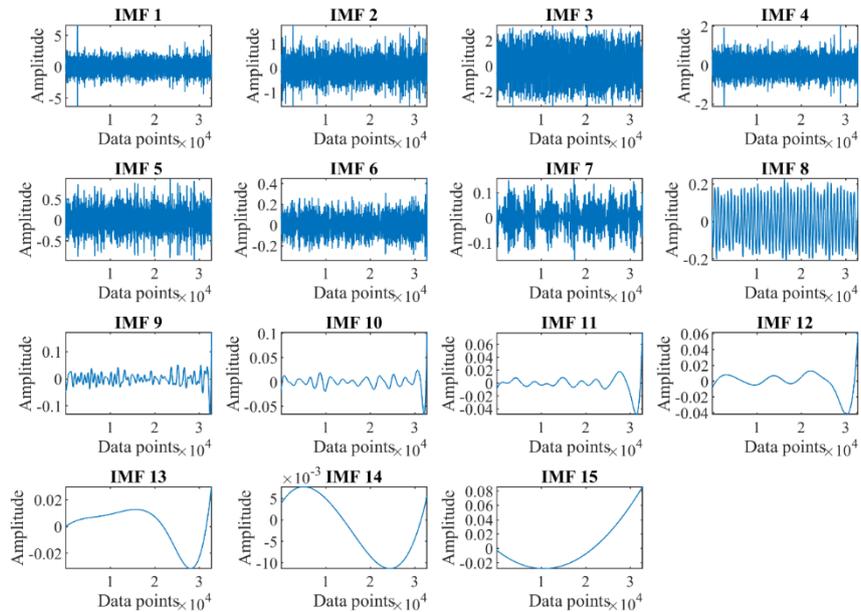

Fig. 54. IMFs generated by applying the EEMD method to the signal shown in Fig. 50

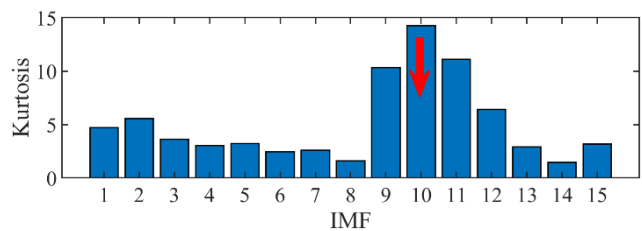

Fig. 55. Kurtosis of the IMFs shown in the Fig. 54

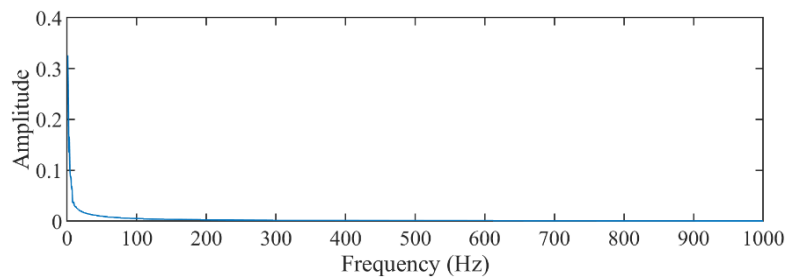

Fig. 56. Envelope spectrum of the IMF 10 (showed in Fig. 54)

### 5.3.3 Diagnosis Results of the CEEMD-based method

The IMF's generated by applying the CEEMD method to the signal shown in **Fig. 50** are shown in Fig. 57. The kurtosis of IMFs shown in **Fig. 57** is displayed in **Fig. 58**. The IMF 10 possesses the greatest kurtosis value. Consequently, the envelope spectrum of IMF 10 is computed and displayed in **Fig. 59**. No peak is found at the bearing defect frequency. This shows that the performance of the CEEMD-based method in this instance is not good.

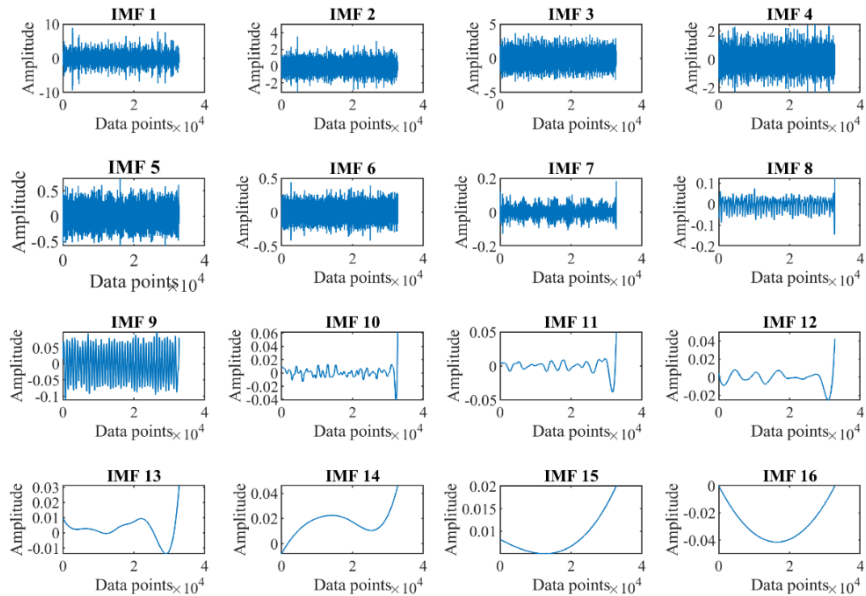

*Fig. 57. IMFs generated by applying the CEEMD method to the signal shown in Fig. 50*

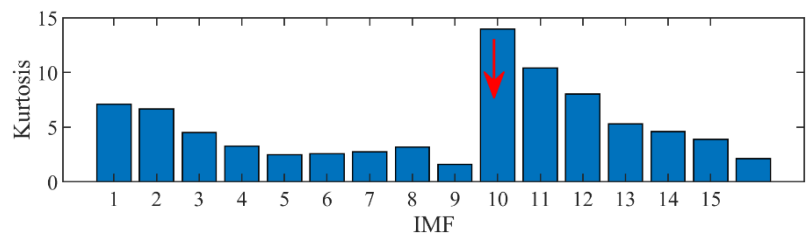

*Fig. 58. Kurtosis of the IMFs shown in Fig. 57*

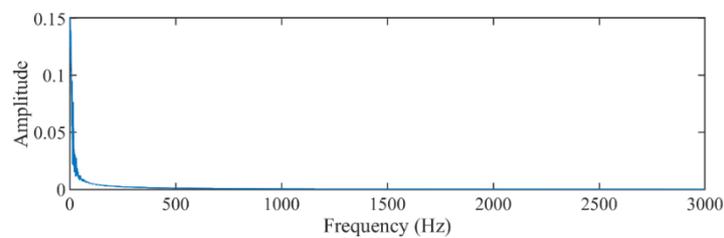

Fig. 59. Envelope spectrum of the IMF 10 *(showed in Fig. 57)*

## 5.3.4 Diagnosis Result of CEEMDAN-based method

The IMF's generated by applying the CEEMDAN method to the signal shown in **Fig. 50** are shown in **Fig. 60**. The kurtosis of IMFs shown in **Fig. 60** is displayed in **Fig. 61**. The IMF-5 possesses the greatest kurtosis value. Consequently, the envelope spectrum of IMF 5 is computed and displayed in **Fig. 62**. A peak is found at the bearing defect frequency. This shows that the performance of the CEEMDAN-based method is satisfactory.

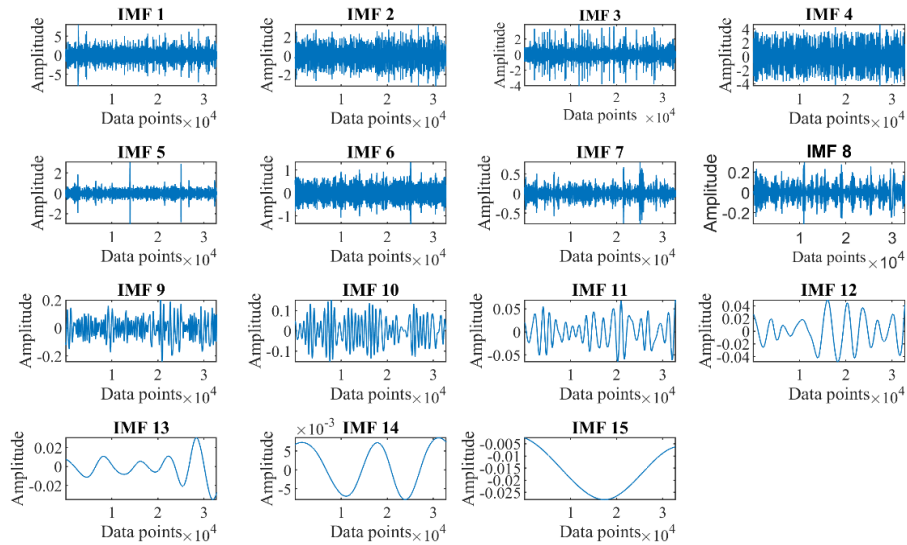

*Fig. 60. IMFs generated by applying the CEEMDAN method to the signal shown in Fig. 50*

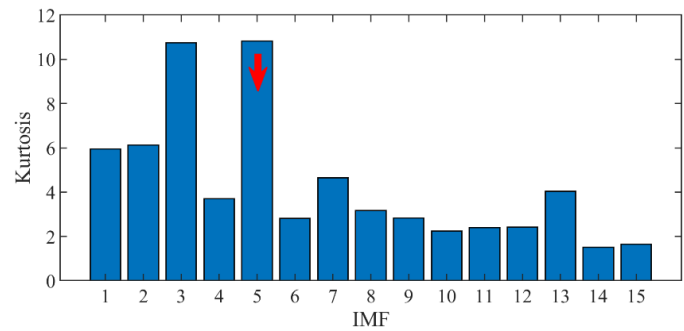

*Fig. 61. Kurtosis of the IMFs shown in Fig. 60*

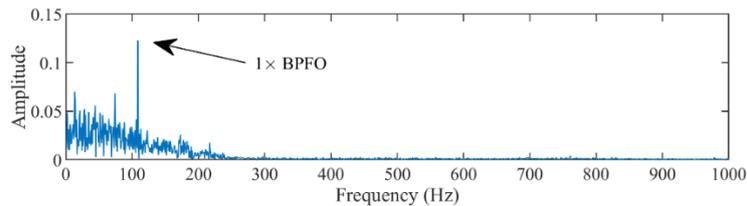

*Fig. 62. Envelope spectrum of the IMF 5 (showed in Fig. 60)*

## 5.4 Identification of bearing defect in the axial piston pump

The axial piston pump is an extremely complicated piece of hydromechanical equipment which has nine pistons, each of which, produces strong deterministic vibration in the machine at the ninth harmonic of shaft speed. This situation of the bearing gives a great challenge when subjected for identifying bearing defects. In the presence of deterministic vibration, it is difficult to extract defect-related information, so overcoming this challenge can be a difficult task. The graphical and schematic structure of the axial piston pump under study is shown in **Fig. 63**. The apparatus used in the test is shown in **Fig. 64**.

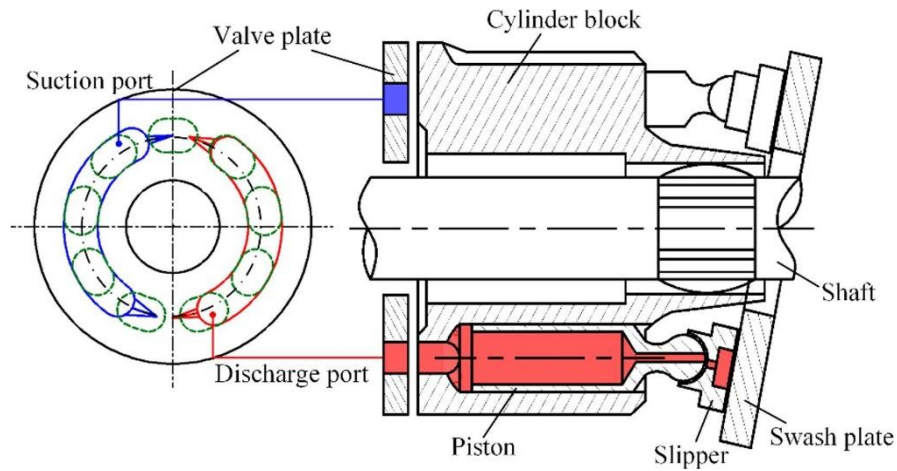

*Fig. 63 Schematic representation of an axial piston pump under study*

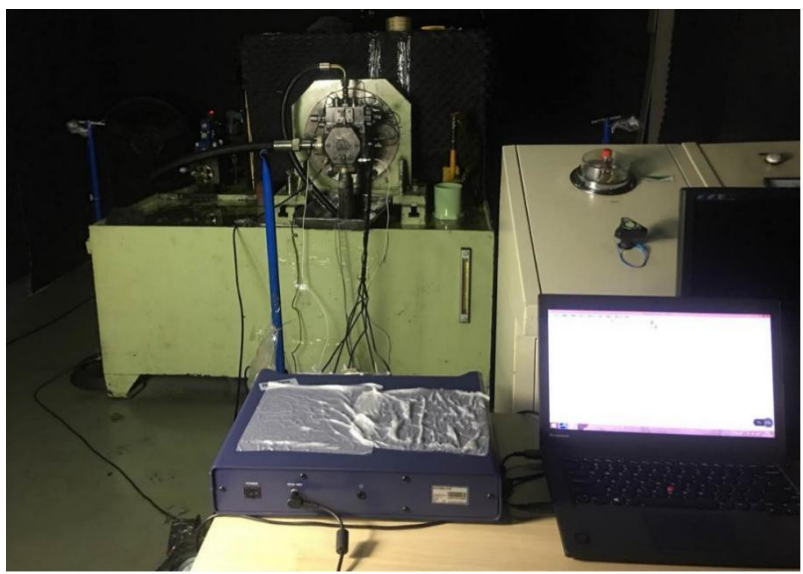

*Fig. 64. Test rig for the axia piston pump*

## 5.4.1 Diagnosis Results of the proposed NPCEEMD-based method

Fig. 65 is the raw vibrational signal collected from the axial pump shown in the **Fig. 64**. The axial pump has an outer race defect in its bearing. **Fig. 66** shows the IMFs obtained from the vibrational signal shown in **Fig. 65**. The mutual information of the IMFs (**Fig. 66**) to the raw signal (**Fig. 65**) is shown in **Fig. 67**. The IMF1, IMF3, and IMF4 have mutual information value greater than 0.1. These IMFs are combined and their envelope spectra are computed and shown in **Fig. 68**. A high peak can be seen at the bearing pass frequency outer race (BPFO). Peaks at the harmonic of BPFO can also be seen in **Fig. 68**.

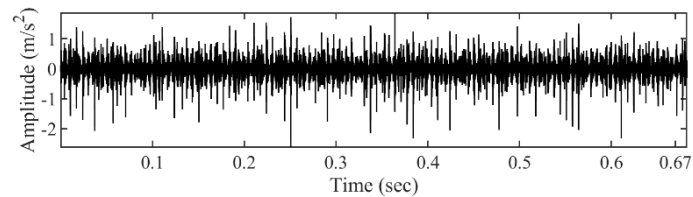

*Fig. 65 Raw signal of the outer race defect condition*

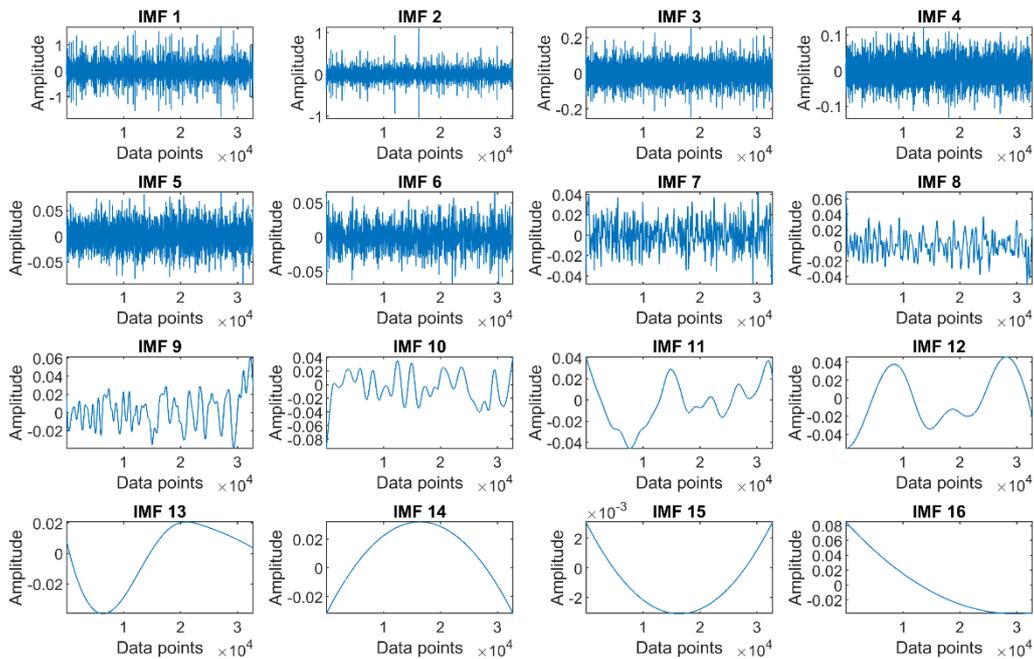

*Fig. 66. IMFs of the signal generated by applying the NPCEEMD method to the signal shown in Fig. 65*

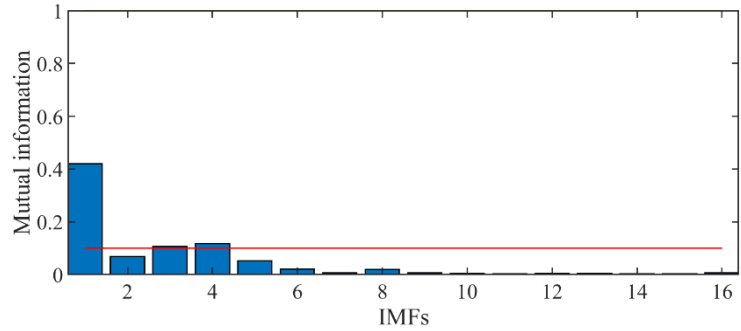

*Fig. 67. Mutual information of IMFs (Fig. 66) to the raw signal (Fig. 65)*

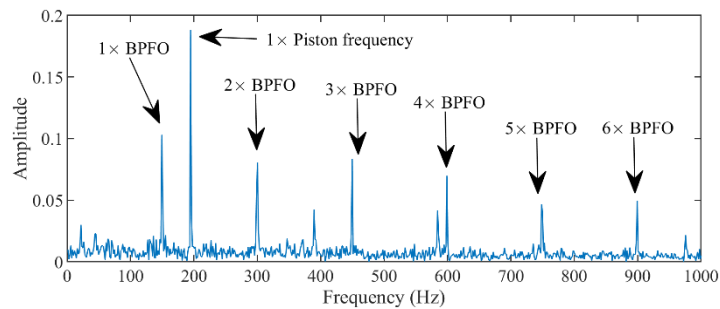

*Fig. 68. Envelope spectrum of the combined signal generated using IMF1, IMF3, and IMF4*

### 5.4.2 Diagnosis Results of EEMD-based method

The IMF's generated by applying the EEMD method to the signal shown in **Fig. 65** are shown in **Fig. 69**. The kurtosis of the IMF shown in **Fig. 69** is given in **Fig. 70**. The IMF 2 has the highest kurtosis value. Therefore, the envelope spectrum of IMF 2 is computed and shown in **Fig. 71**. The peak can be found at the bearing defect frequency. In this particular case, the performance of the EEMD method is almost identical with the proposed NPCEEMD method.

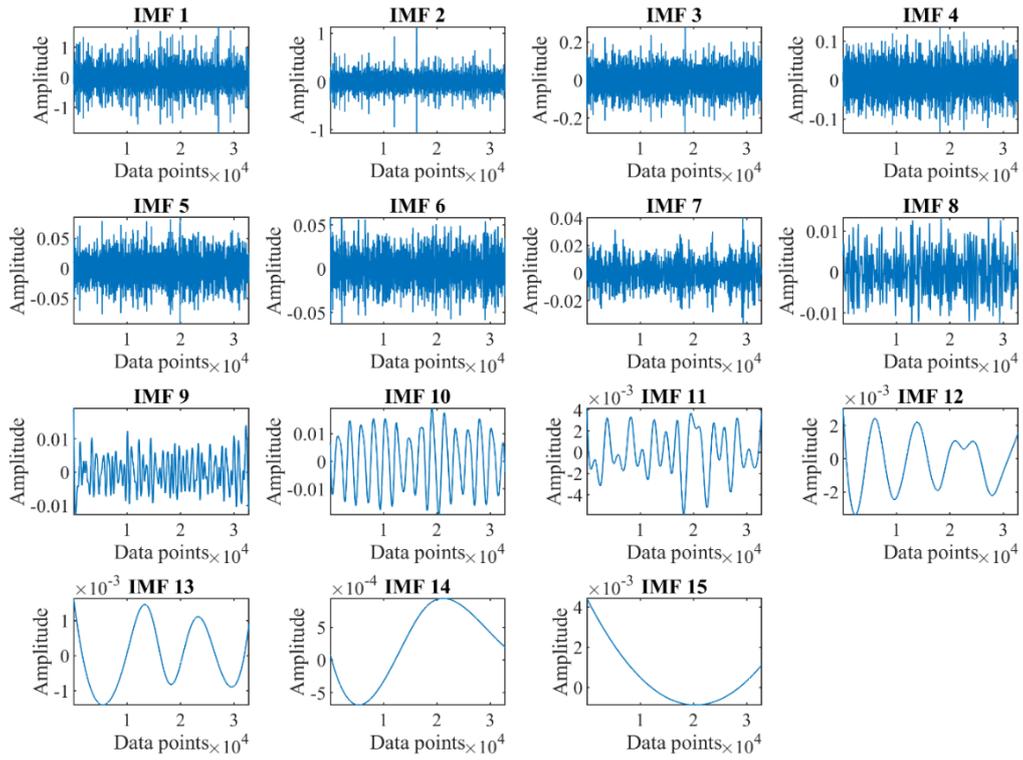

*Fig. 69. IMFs generated by applying the EEMD method to the signal shown in Fig. 65*

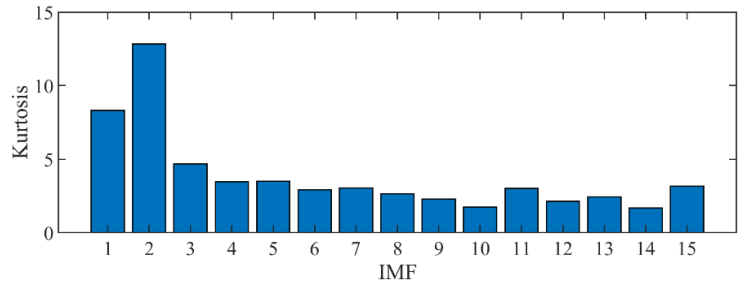

*Fig. 70. Kurtosis of the IMFs showed in Fig. 69*

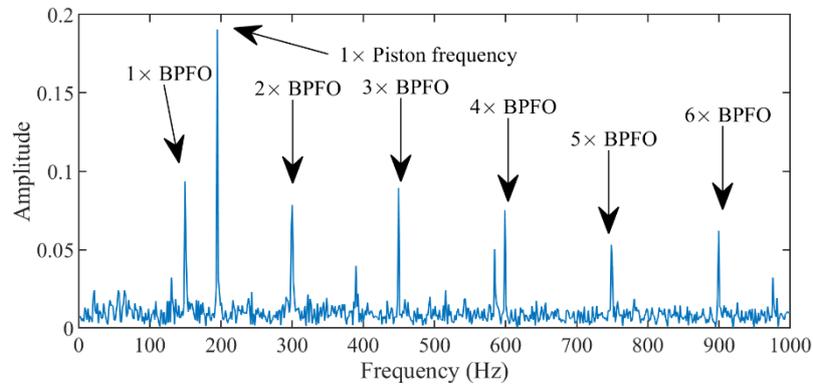

*Fig. 71. Envelope spectrum of the IMF2 (showed in Fig. 69)*

## 5.4.3 Diagnosis Results of CEEMD-based method

The IMF's generated by applying the CEEMD method to the signal shown in **Fig. 65** are shown in **Fig. 72**. The kurtosis of the IMF shown in **Fig. 72** is displayed in **Fig. 73**. The IMF 2 has the highest kurtosis. Therefore, the envelope spectrum of IMF 2 is computed and shown in **Fig. 74**. The peak can be found at the bearing defect frequency. In this particular case, the performance of the CEEMD-based method is almost identical with the proposed NPCEEMD method.

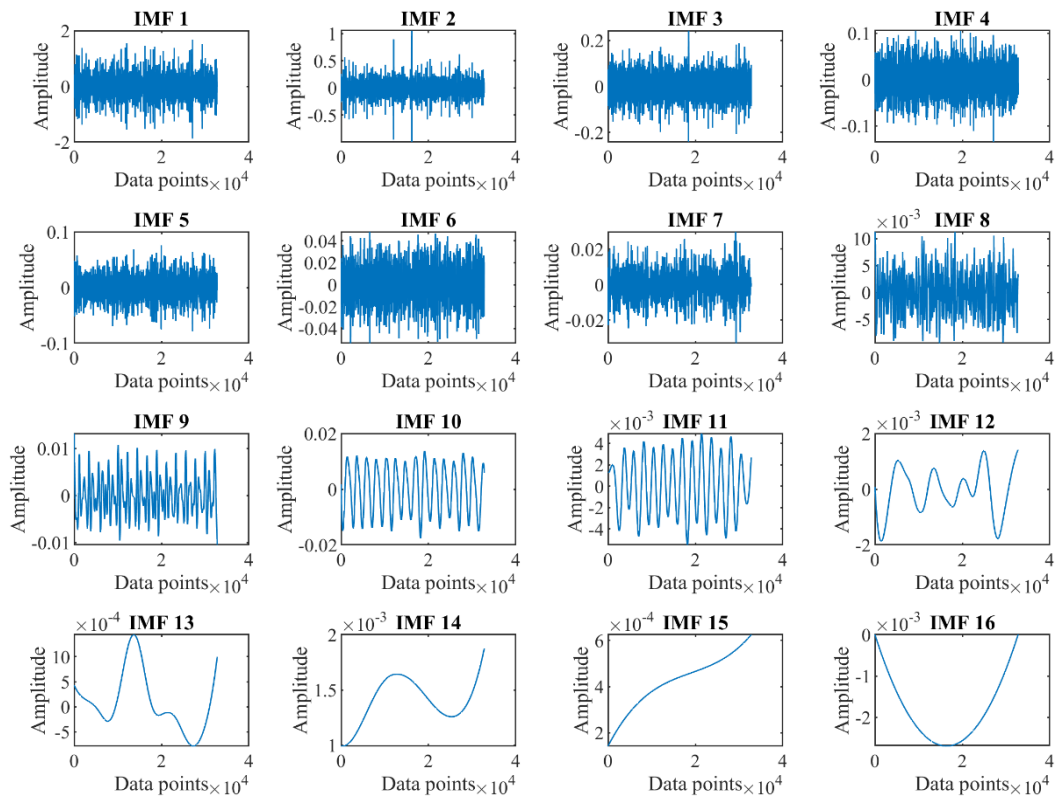

*Fig. 72. IMFs generated by applying the CEEMD method to the signals shown in Fig. 65*

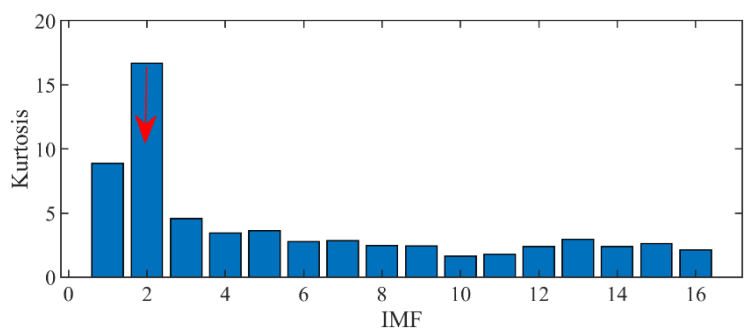

*Fig. 73. Kurtosis of the IMFs showed in Fig. 72*

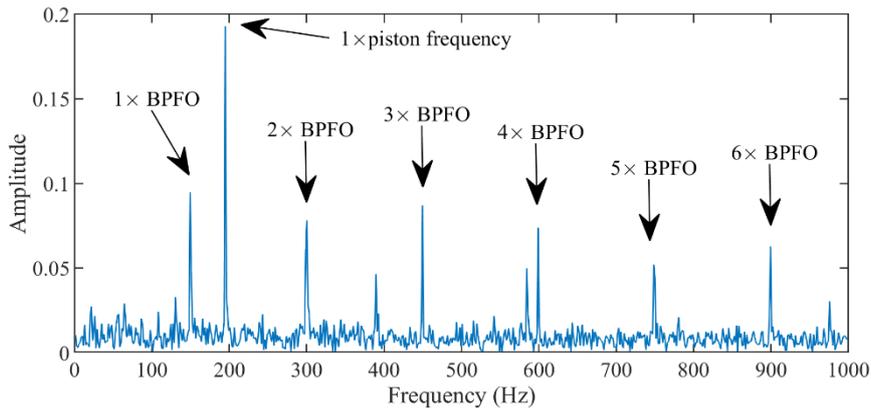

*Fig. 74. Envelope spectrum of the IMF2 (showed in Fig. 72)*

### 5.4.4 Diagnosis Results of CEEMDAN-based method

The IMF's generated by applying the CEEMD method to the signal shown in **Fig. 65** are shown in **Fig. 75**. The kurtosis of the IMF shown in **Fig. 75** is displayed in **Fig. 76**. The IMF 2 has the highest kurtosis value. Therefore, the envelope spectrum of IMF 2 is computed and shown in **Fig. 77**. The peak can be found at the bearing defect frequency. In this particular case, the performance of the CEEMDAN-based method is found to be satisfactory.

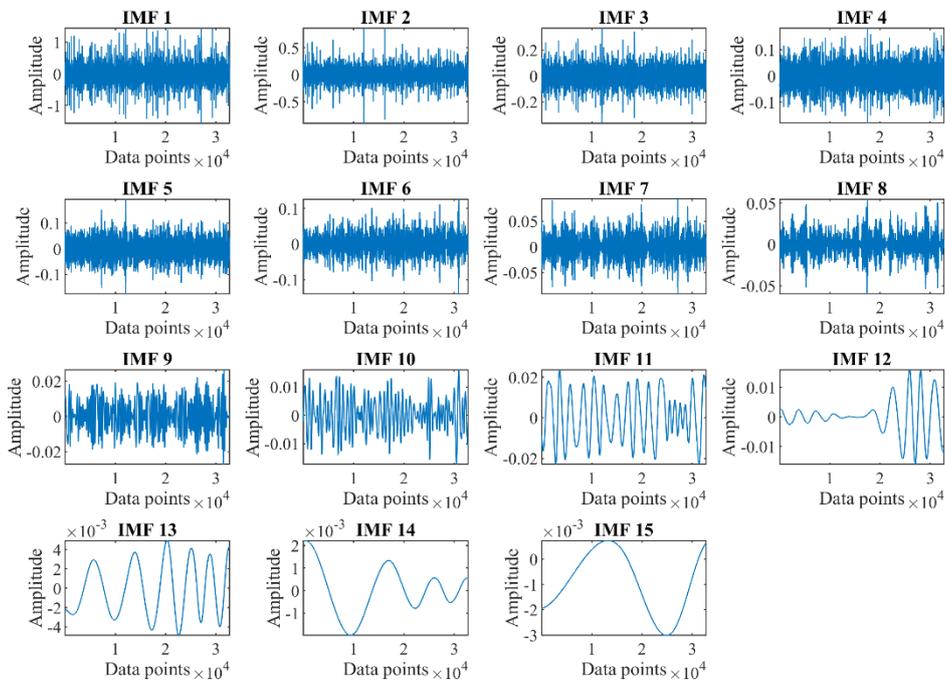

*Fig. 75. IMFs generated by applying the CEEMDAN method to the signals shown in Fig. 65*

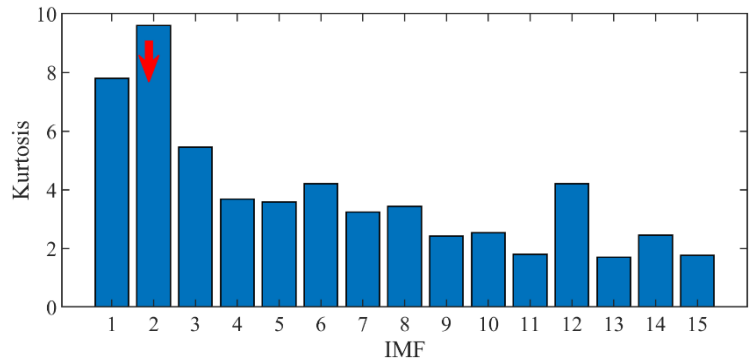

*Fig. 76. Kurtosis of the IMFs showed in Fig. 75*

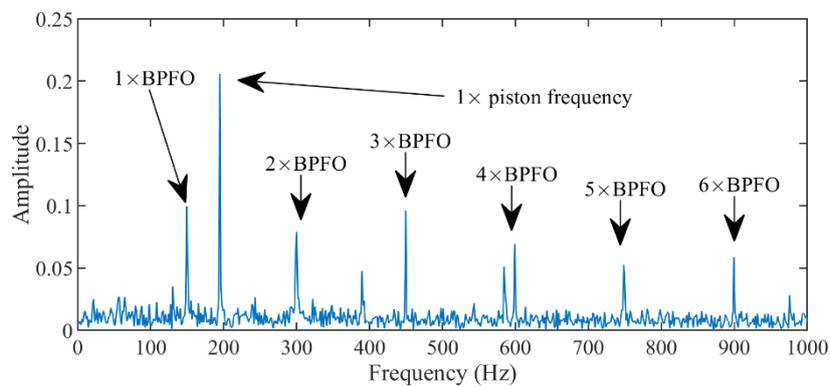

*Fig. 77. Envelope spectrum of the IMF2 (showed in Fig. 75)*

**5.5 Identification of bearing defect using acoustic data**

The image shown in **Fig. 78** represents the rolling element bearing test setup. At either end of the shaft, two bearings provide support (bearing 1 and bearing 2). The shaft receives its power from an AC motor rated at 346 watts. A step-pulley configuration is used to transfer the power from the motor to the shaft. A disc weighting 2 kilograms is fastened to the shaft at the centre of the shaft, in between bearings 1 and 2, and it rotates together with the shaft. A microphone was used to acquire the acoustic data, as shown in **Fig. 78**. The acoustic data set so obtained has been made available publicly [39]. A sampling rate of 70,000 samples per second was used to record the vibration data.

Cylindrical roller bearings are used in the test (NBC: NU205E). The rectangular grooves were produced using an electric discharge machining (EDM) process to simulate the effect of a localized defect. The experiment was performed at an operating frequency of 2050 revolutions per minute.

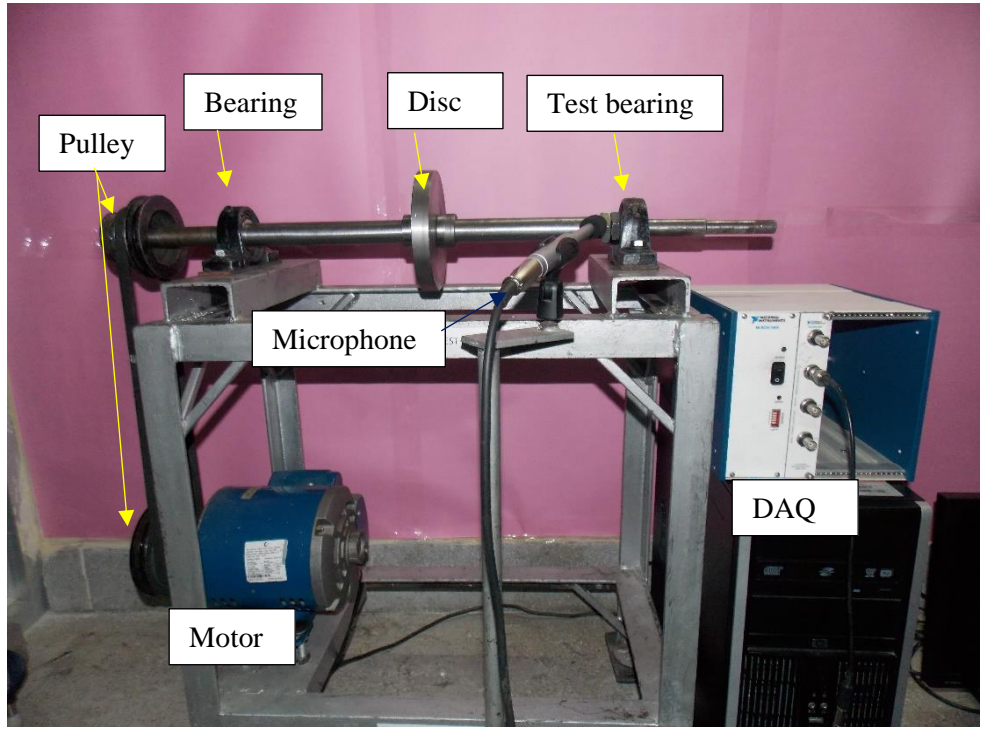

*Fig. 78. A typical image of the testing apparatus*

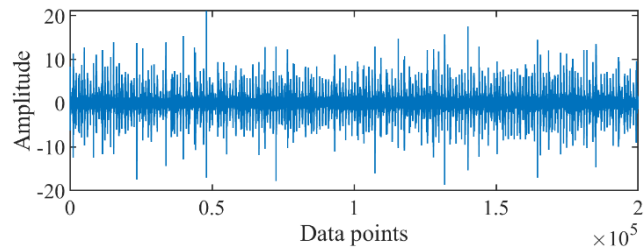

*Fig. 79 Raw signal of inner race defect condition*

### 5.5.1 Diagnosis Results of the proposed NPCEEMD-based method

**Fig. 79** is the raw vibrational signal collected from the test rig shown in **Fig. 78**. This data set is large, with 200,000 samples. **Fig. 80** shows the IMFs obtained from the vibrational signal shown in **Fig. 79**. The mutual information of the IMFs (**Fig. 80**) to the raw signal (**Fig. 79**) is shown in **Fig. 81**. The IMF1, IMF2, and IMF3 have mutual information value greater than 0.1. These IMFs are combined, and their envelope spectra are computed and shown in **Fig. 82**. This spectrum has the highest peak at the second harmonics of shaft speed (2X) and a significant peek at the BPFI. This confirms the presence of an inner-race defect. A sideband of the rotation frequency can be seen around BPFI, which indicates that there is a modulation effect due to the shaft rotation frequency.

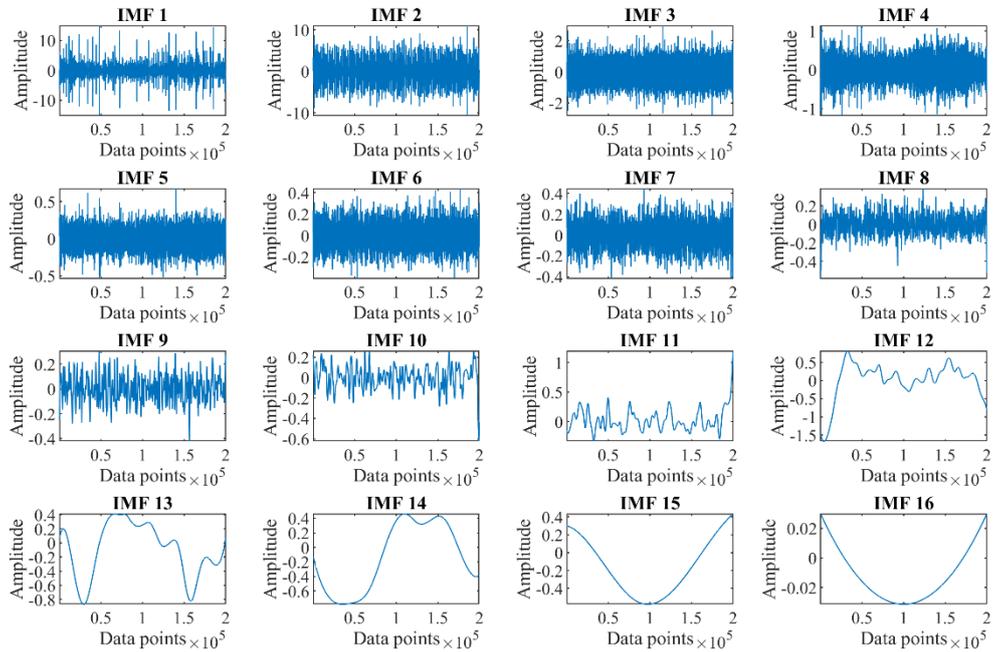

*Fig. 80. IMFs generated by applying the NPCEEMD method to signal shown in Fig. 79*

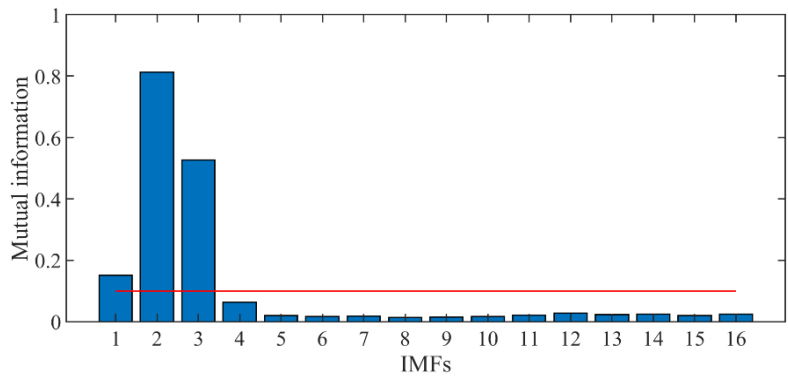

*Fig. 81. Mutual information of IMFs (Fig. 80) to the raw signal (Fig. 79)*

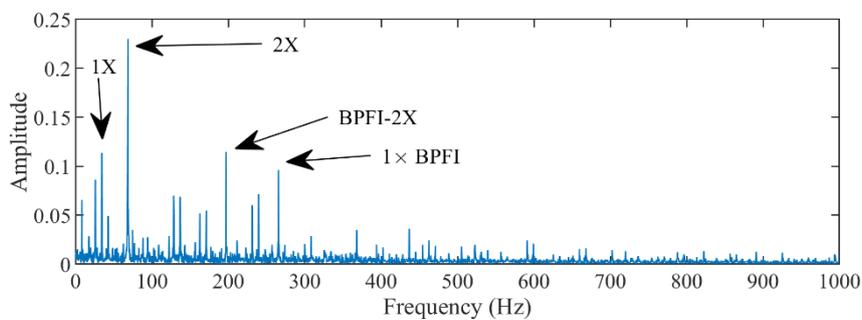

*Fig. 82. Envelope spectrum of the combined signal generated using IMF1, IMF2, and IMF3*

## 5.5.2 Diagnosis Results of EEMD-based method

The IMF's generated by applying the EEMD method to the signal shown in **Fig. 79** are shown in **Fig. 83**. The kurtosis of the IMF shown in Fig. 83 is displayed in **Fig. 84**. The IMF 1 has the highest kurtosis value. Therefore, the envelope spectrum of IMF 1 is computed and shown in **Fig. 85**. The peak can be found at the bearing defect frequency. In this particular case, the accuracy of the EEMD-based method is almost identical with the proposed NPCEEMD method.

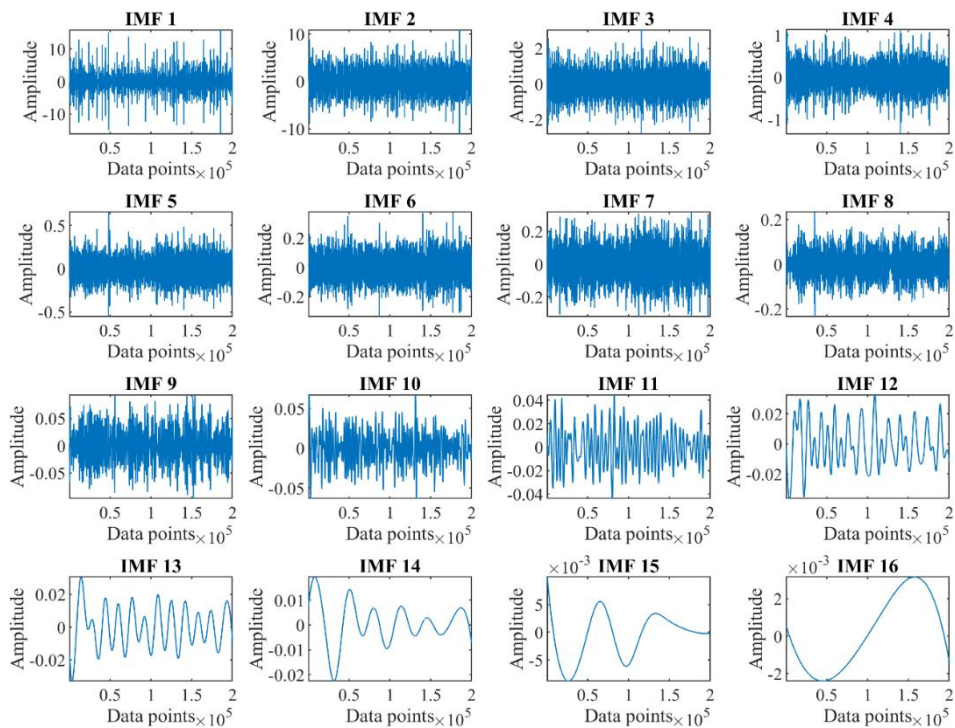

*Fig. 83. IMFs generated by applying the EEMD method to the signal shown in Fig. 79*

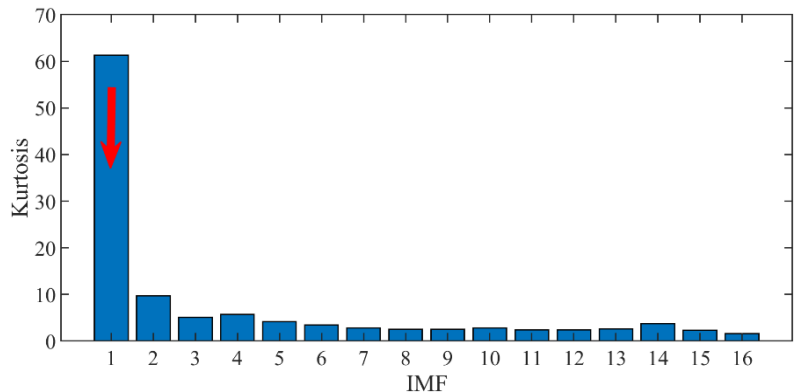

*Fig. 84. Kurtosis of IMFs showed in the Fig. 83*

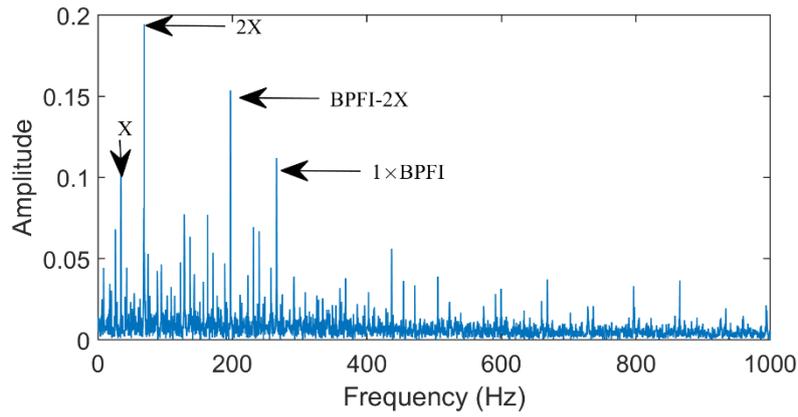

*Fig. 85. Envelope spectrum of the IMF1(showed in Fig. 83)*

### 5.5.3 Diagnosis Results of CEEMD-based method

The IMF's generated by applying the CEEMD method to the signal shown in **Fig. 79** are shown in **Fig. 86**. The kurtosis of the IMF shown in **Fig. 8**6 is displayed in **Fig. 87.** The IMF 1 has the highest kurtosis value. Therefore, the envelope spectrum of the IMF 1 is computed and shown in **Fig. 88**. The peak can be found at the bearing defect frequency.

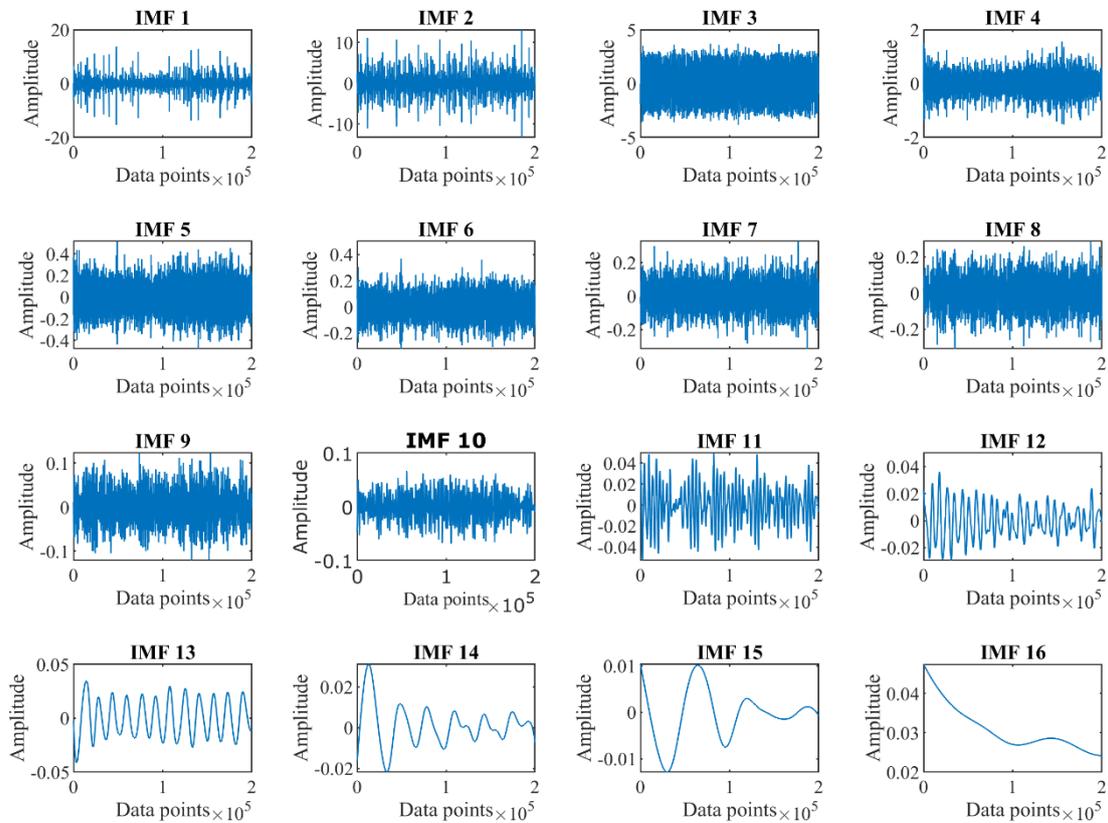

*Fig. 86. IMFs generated by applying the CEEMD method to the signal shown in Fig. 79*

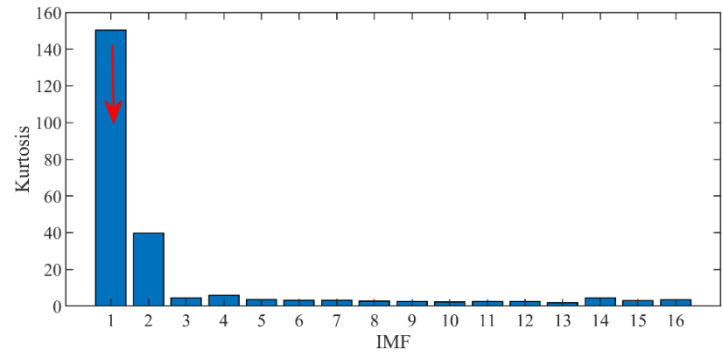

*Fig. 87. Kurtosis of the IMFs showed in Fig. 86*

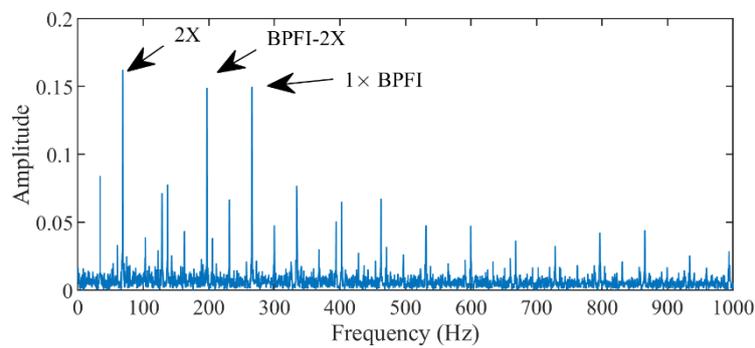

*Fig. 88. Envelope spectrum of the IMF1(showed in Fig. 86)*

### 5.5.4 Diagnosis Results of CEEMDAN-based method

The IMF's generated by applying the CEEMDAN method to the signal shown in **Fig. 79** are shown in **Fig. 89**. The kurtosis of the IMF shown in **Fig. 89** is displayed in **Fig. 90**. The IMF 1 has the highest kurtosis value. Therefore, the envelope spectrum of the IMF 1 is computed and shown in **Fig. 91**. The peak can be found at the bearing defect frequency.

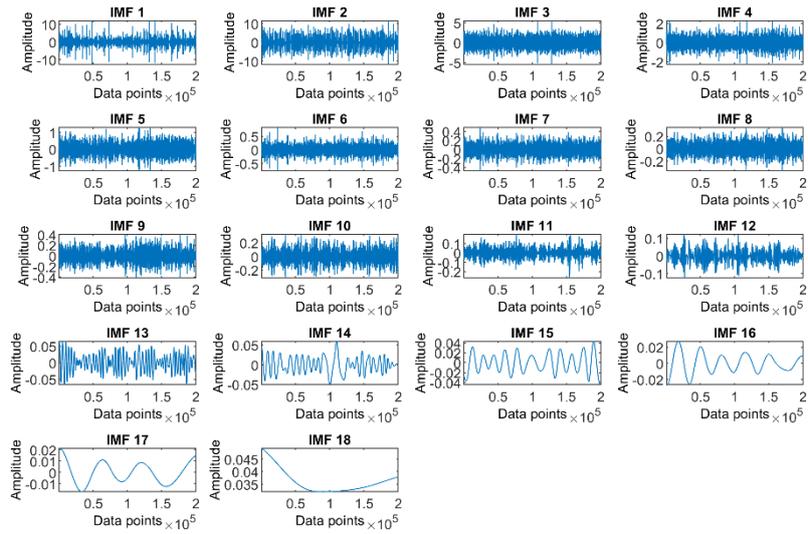

*Fig. 89. IMFs generated by applying the CEEMDAN method to the signal shown in Fig. 79*

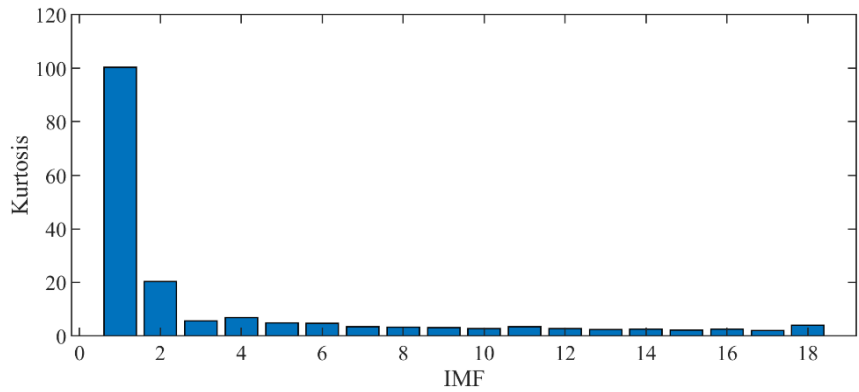

*Fig. 90. Kurtosis of IMFs showed in Fig. 89*

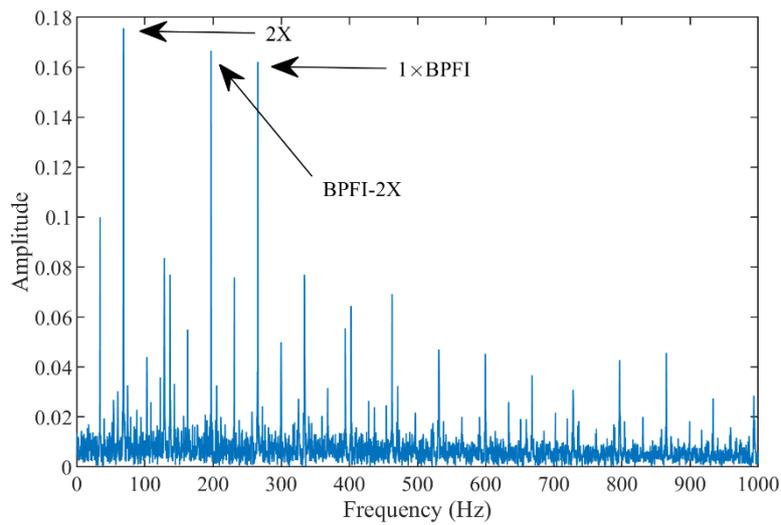

*Fig. 91. Envelope spectrum of the IMF1(showed in Fig. 89)*

The above-carried experimental analysis on the artificially seeded fault summarizes that the existing EEMD and CEEMD-based methods have been found to perform well. On the other hand, while identifying faults that are created using natural progressive wear, such as in the case of IMS data and the XJTU data set, identification of bearing defect could not be performed by applying the existing EEMD-based method. The CEEMD-based technique could identify bearing defects for the IMS data set, but failed to identify bearing defects for the XJTU data set. The CEEMDAN-based method could identify bearing defects for the XJTU data set but failed to identify bearing defects for the IMS data set and simulative data of degraded bearing. However, our proposed NPCEEMD-based method has been found capable to identify bearing defects for defects for the IMS as well as XJTU data set.

## 6. Conclusion

A non-parametric ensemble empirical mode decomposition (NPCEEMD)-based decomposition is proposed for extracting weak defect features to identify bearing defects in rotary machinery. The proposed NPCEEMD-based defect identification methodology has been successfully tested on the axial piston pump and acoustic data set (generated from the cylindrical roller bearing), and accelerated degradation data of IMS centre and Xi'an Jiatong University. The following conclusions can be drawn from the study:

- The proposed NPCEEMD method is non-parametric in nature, which means that, it does not require the selection of an efficient SNR and many ensemble trials for each signal processing step. By applying the NPCEEMD-based method, a satisfactory correlation between IMFs and the raw signal can be achieved with a less number of ensemble operation. In other words, only a few ensemble trials are sufficient to ensure effective decomposition. As a result, this method become computational efficient.

- In the analysis discussed in **Appendix A.2**, it is concluded that the decomposition performance of the existing methods such as EEMD and CEEMD is substantial at ensemble number 100 but extremely poor at ensemble number 10. On the other hand, our proposed NPCEEMD-based method performs well even for a smaller ensemble size of 10. This investigation indicates that any changes in NPCEEMD parameters have a little effect on their decomposition performance.

- A criterion based on mutual information between the IMFs and the raw signal is proposed for selecting the sensitive IMFs. We have employed the "nearest neighbors approach" to compute the mutual information between the raw signal and IMFs,

- generated from the NPCEEMD method. The IMFs with mutual information greater than 0.1 are selected and combined to form a resulting IMF. Envelope demodulation is applied to resulting IMF for the identification of defects.

- It is concluded that the IMFs with mutual information value less than 0.1 have no significance and can be ignored. Hence, the proposed NPCEEMD based method is an appropriate tool for selecting the IMFs with defect related information and filtering out the remaining ones.

- A comparison of the proposed NPCEEMD-based approach has been done with the existing decomposition methods (EEMD, CEEMD, and CEEMDAN). The comparison indicates that that the existing methodology can perform effectively on the artificially seeded defect. However, the defects, generated by natural progressive wear, as in the case of IMS data and the XJTU data set, couldn't be identified using EEMD-based method. The CEEMD-based method could identify the defects from the simulative data set of degraded bearings and the IMS data set, but failed to identify the defects for the XJTU data set. The CEEMDAN-based method could identify defects using the XJTU data set, but failed to identify defects for the IMS data set and simulative data of degraded bearing. The proposed method performed efficiently in every situation.

**Acknowledgement**

The authors are grateful to the support of the Zhejiang Natural Science Foundation of China (No. LD21E050001), Wenzhou Major Science and Technology Innovation Project of China (Nos. ZG2021019, ZG2021027), National Center of Science under Sheng2 (No. UMO-2021/40/Q/ST8/00024) and Ouyue Haizhi Young Talent Plan Program of Wenzhou

(No. 202211301235057486)

## Appendix A.1: Effect of hurst exponent on the NPCEEMD performance

To study the effect of Hurst exponent on decomposition performance of the NPCEEMD method, we need to compute the IMFs using different values of H. IMFs of noisy signal (Fig. 8) having an SNR of -30 dB are computed using different H values from 0.1 to 0.9, and are shown in Figs. A.1 to A.9. Our findings show that as the Hurst exponent increases from 0.1 to 0.9, mode mixing increase. Therefore, we have chosen an H value of 0.1 for generating noise for carrying out NPCEEMD.

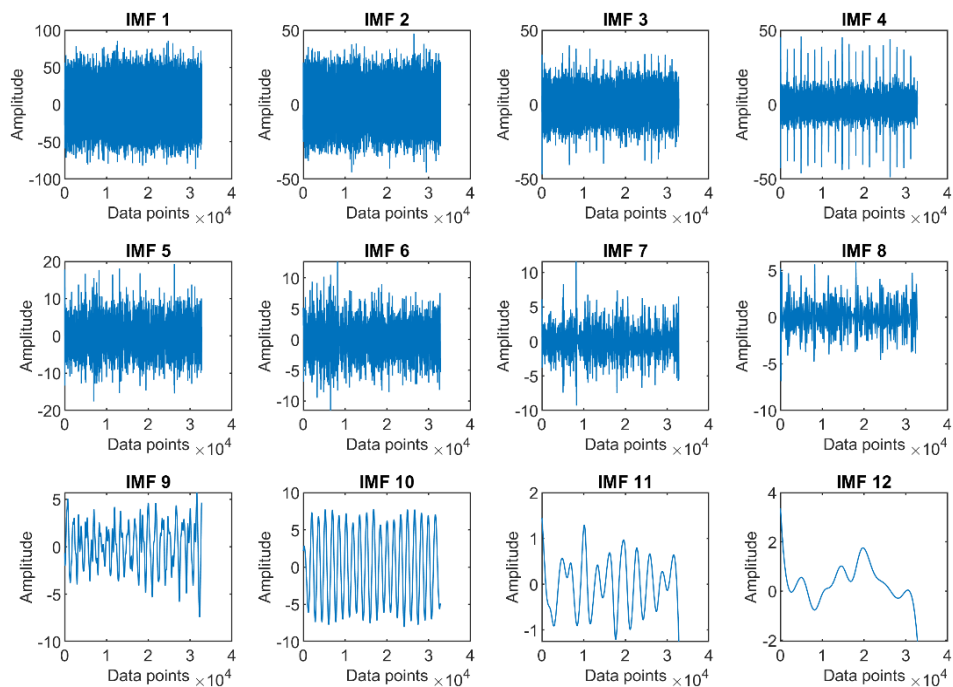

Fig. A1. NPCEEMD results at H equal to 0.1

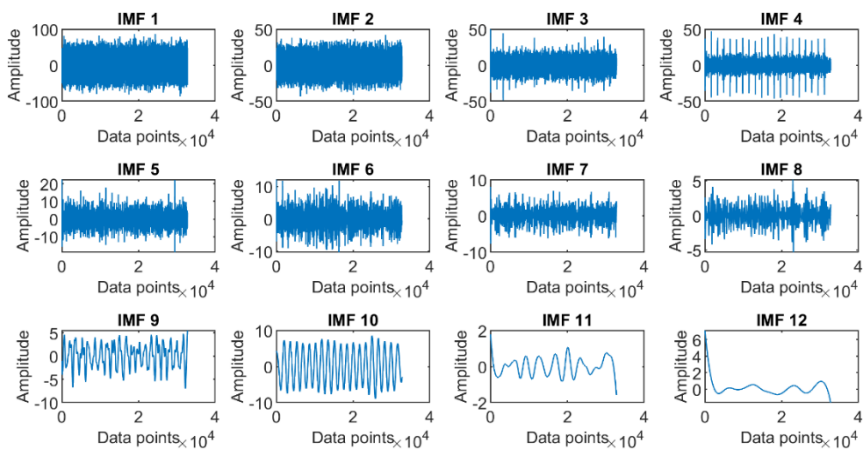

Fig. A2. NPCEEMD results at H equal to 0.2

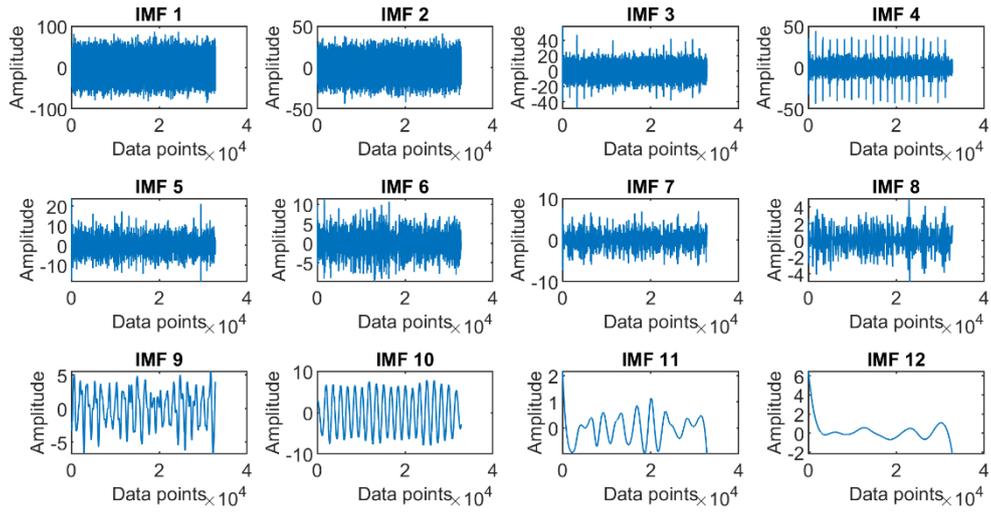

Fig. A3. NPCEEMD results at H equal to 0.3

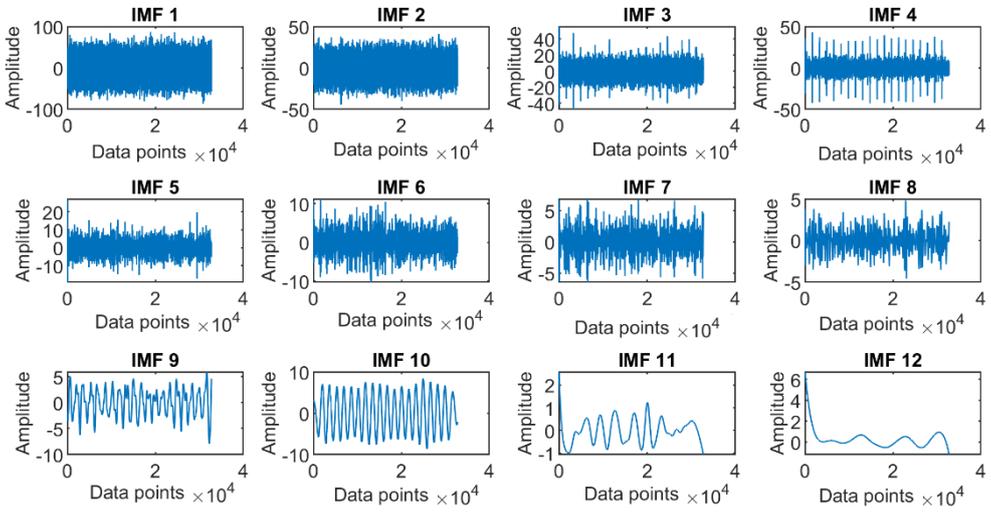

Fig. A4. NPCEEMD results at H equal to 0.4

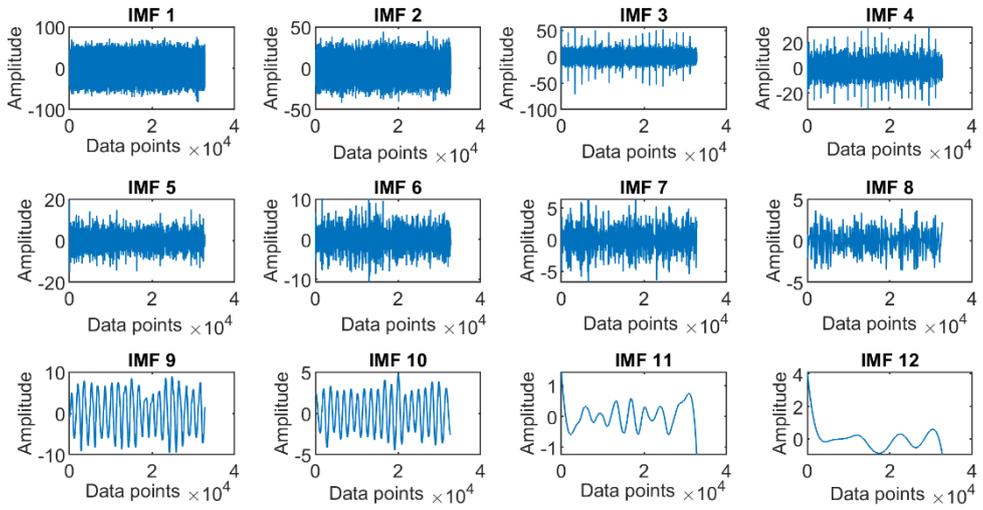

Fig. A5. NPCEEMD results at H equal to 0.5

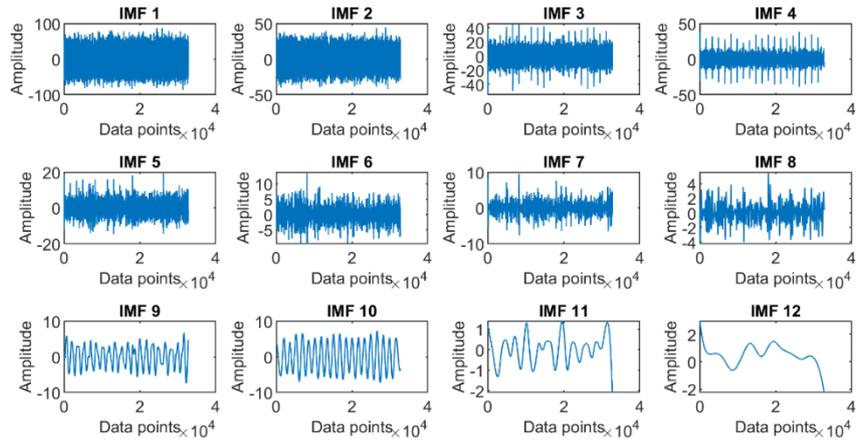

Fig. A6. NPCEEMD results at H equal to 0.6

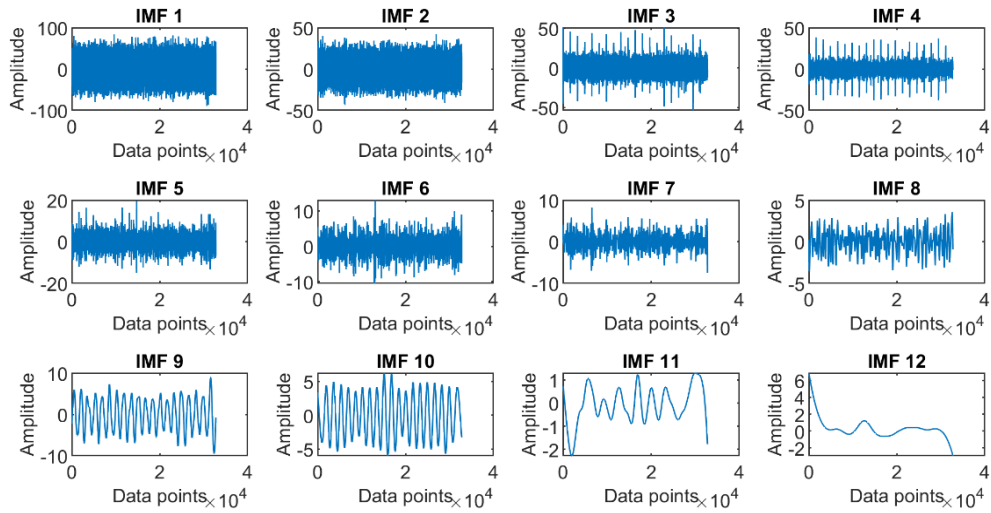

Fig. A7. NPCEEMD results at H equal to 0.7

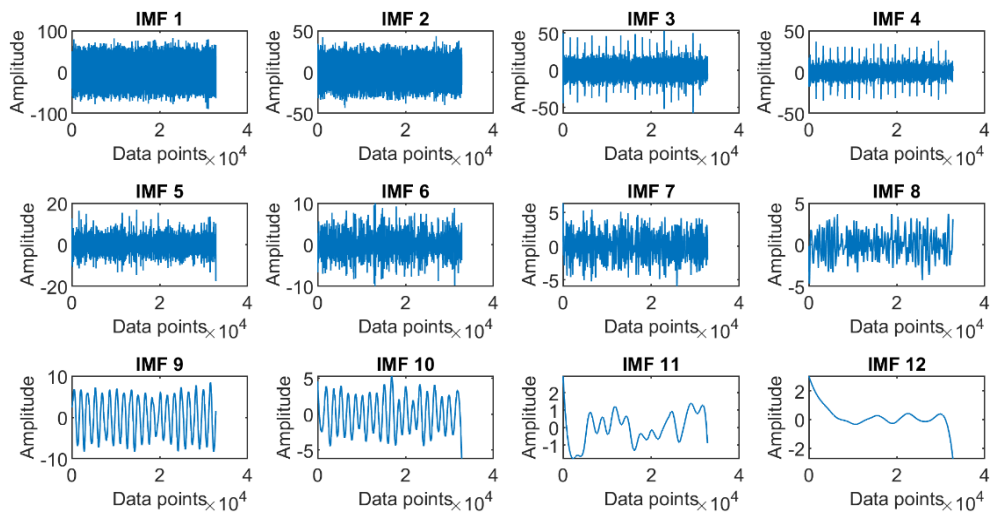

Fig. A8. NPCEEMD results at H equal to 0.8

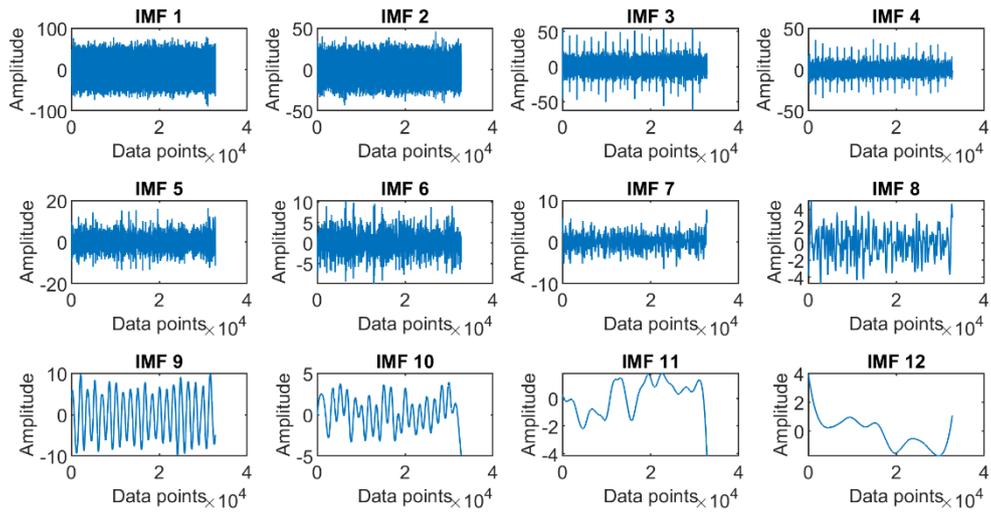

Fig. A9. NPCEEMD results at H equal to 0.9

# Appendix A.2: Effect of ensemble number on the decomposition performance

The effect of modifying parameters on the decomposition performance of EEMD, CEEMD, CEEMDAN and NPCEEMD has been investigated. The results of each approach were computed by applying these algorithms to the simulated data with a signal-to-noise ratio of -30 dB (shown in Fig. 8) using two distinct ensemble number values: 10 and 100. The results of EEMD for ensemble sizes 10 and 100 are depicted in Figs A.10 and A.11, respectively. The results of CEEMD for ensemble sizes 10 and 100 are depicted in Figs. A.12 and A.13, respectively. The results of CEEMDAN for ensemble sizes 10 and 100 are depicted in Figs. A.14 and A.15, respectively. The results of the NPCEEMD for ensemble sizes 10 and 100 are depicted in Figs. A.16 and A.17, respectively. The performance of EEMD and CEEMD is substantial at ensemble number 100, but it is extremely poor at ensemble number 10. The CEEMDAN can extract impulses at both low and high ensemble number, but the mode mixing is more evident in the low frequency part. On the other hand, the NPCEEMD method performs well for extracting low and high frequency signal, at both ensemble sizes 10 and 100. The investigation indicates that any changes in NPCEEMD parameters have little effect on their performance. The demonstration shown in this work proves that the proposed NPCEEMD can achieve superior performance at a low ensemble number, resulting in a reduction in the requirement for computational effort.

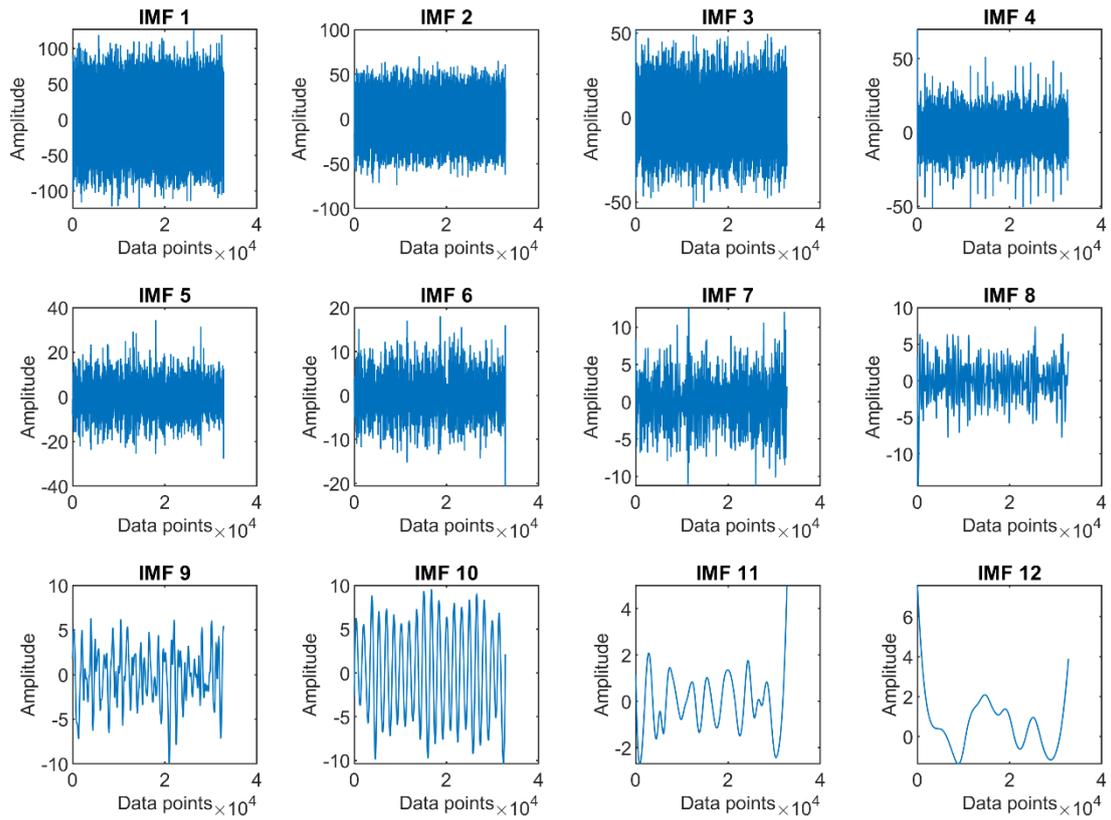

Fig. A.10. IMFs generated by applying EEMD to the signal shown in Fig. 8 using ensemble number equal to 10

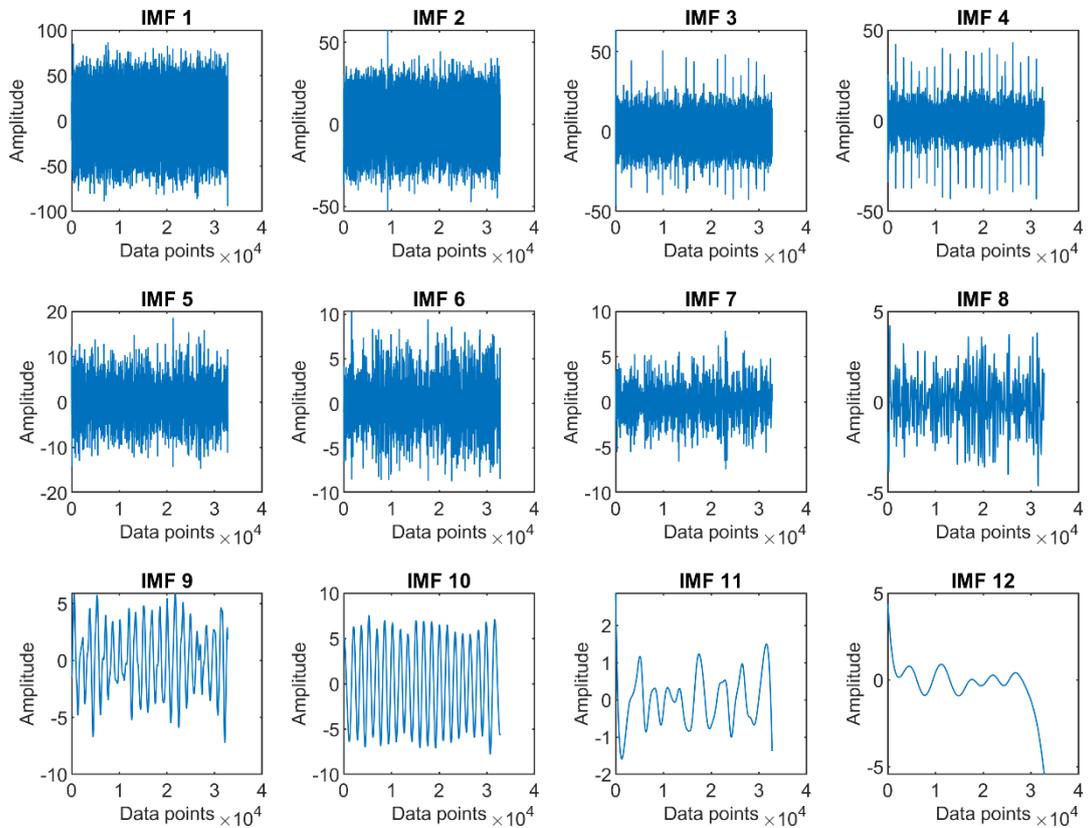

Fig. A11. IMFs generated by applying EEMD to the signal shown in Fig. 8 using ensemble number equal to 100

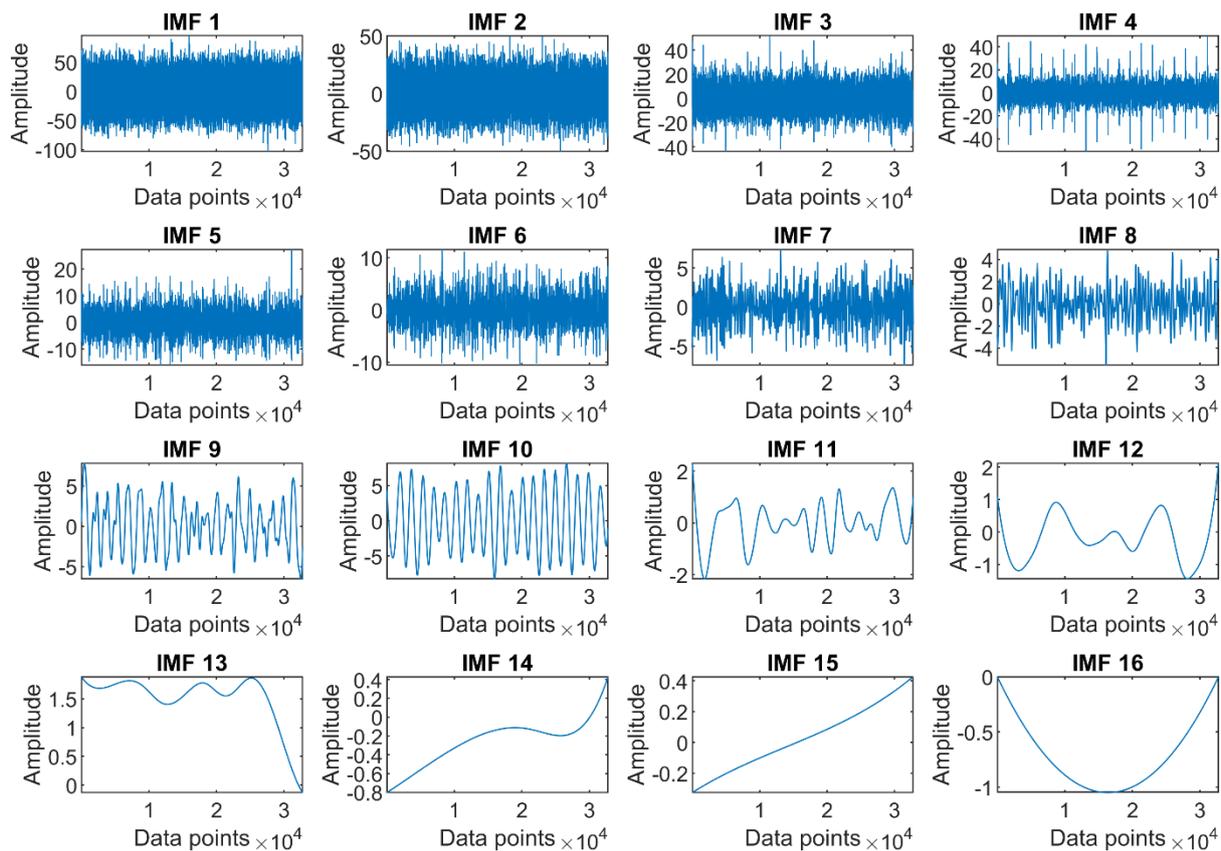

Fig. A12. IMFs generated by applying CEEMD to the signal shown in Fig. 8 using ensemble number equal to 10

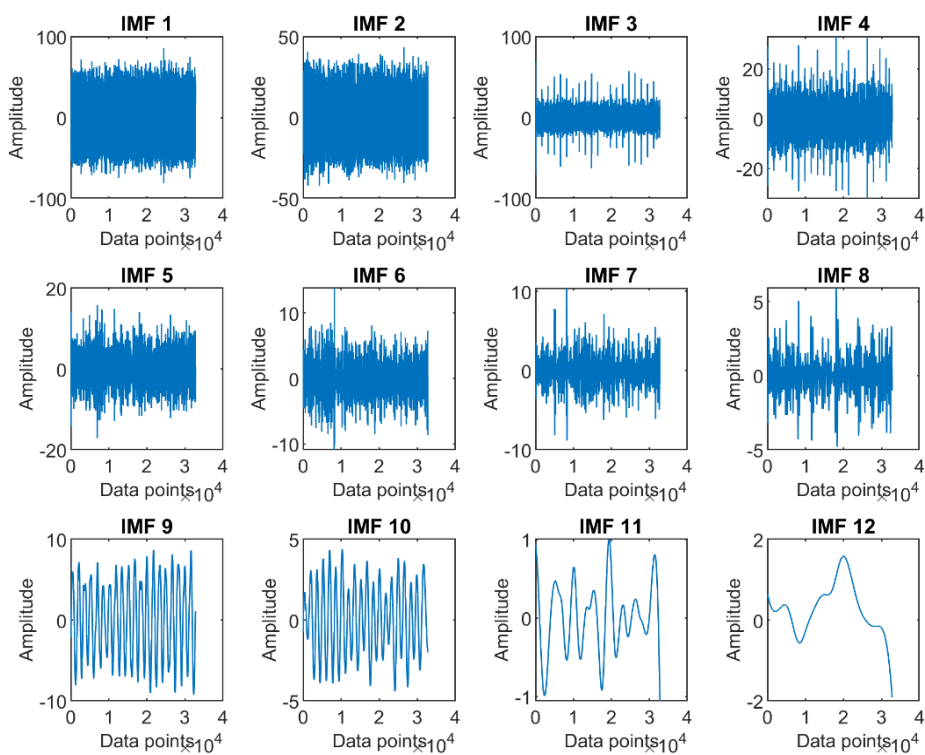

Fig. A13. IMFs generated by applying CEEMD to the signal shown in Fig. 8 using ensemble number equal to 100

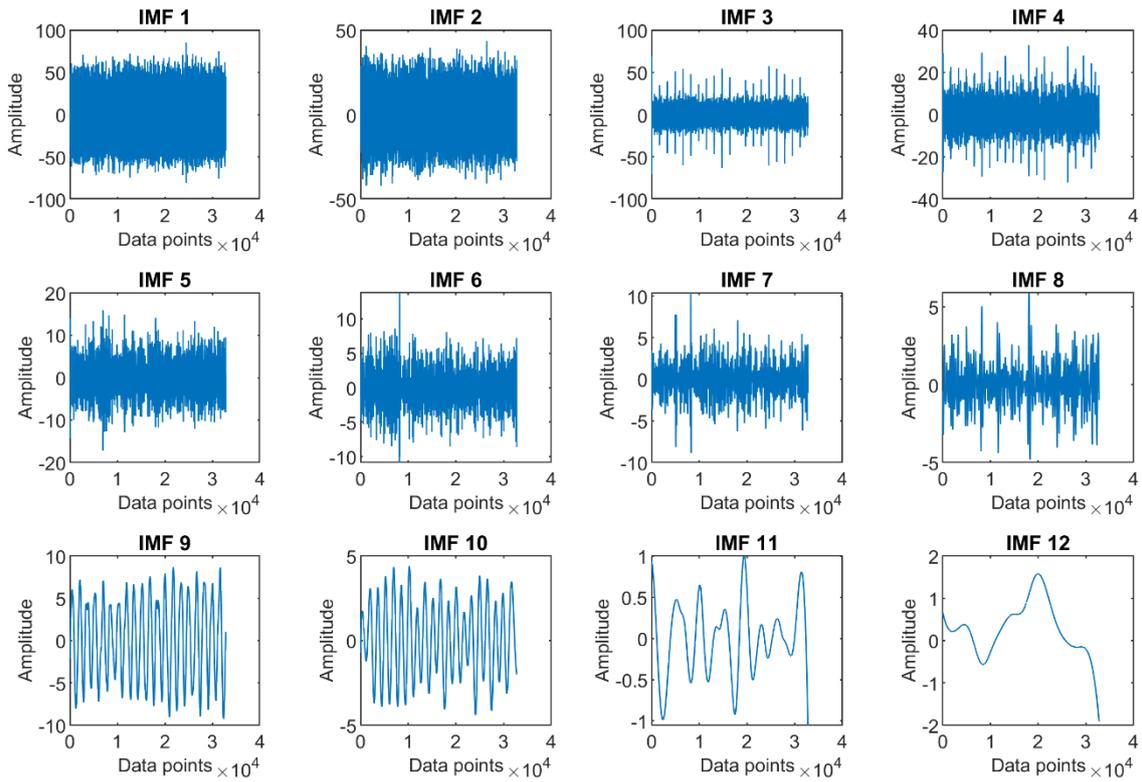

Fig. A14. IMFs generated by applying CEEMDAN to the signal shown in Fig. 8 using ensemble number equal to 10

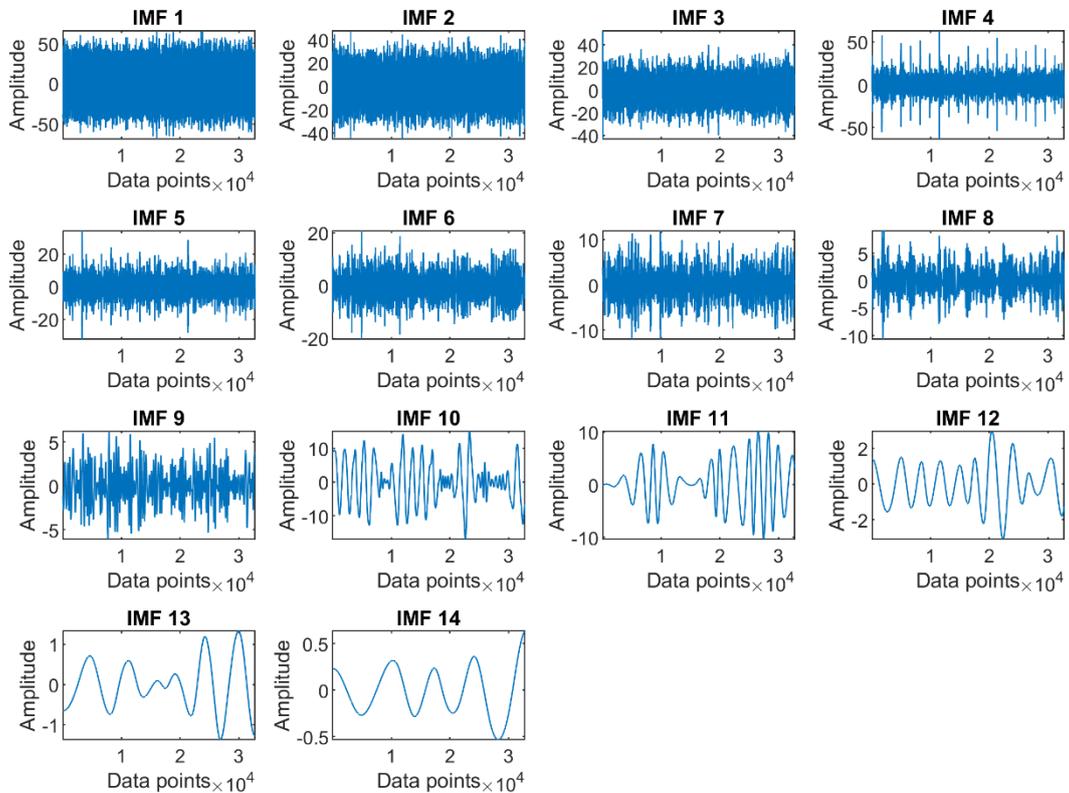

Fig. A15. IMFs generated by applying CEEMDAN to the signal shown in Fig. 8 using ensemble number equal to 100

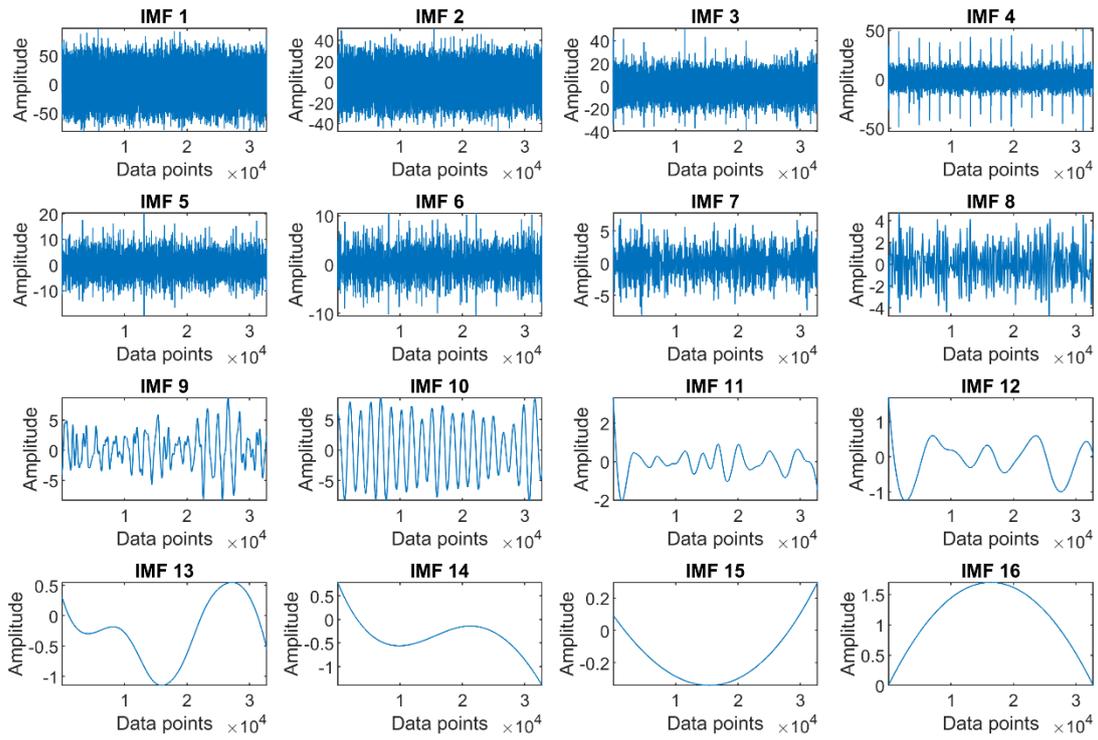

Fig. A16. IMFs generated by applying NPCEEMD to the signal shown in Fig. 8 using ensemble number equal to 10

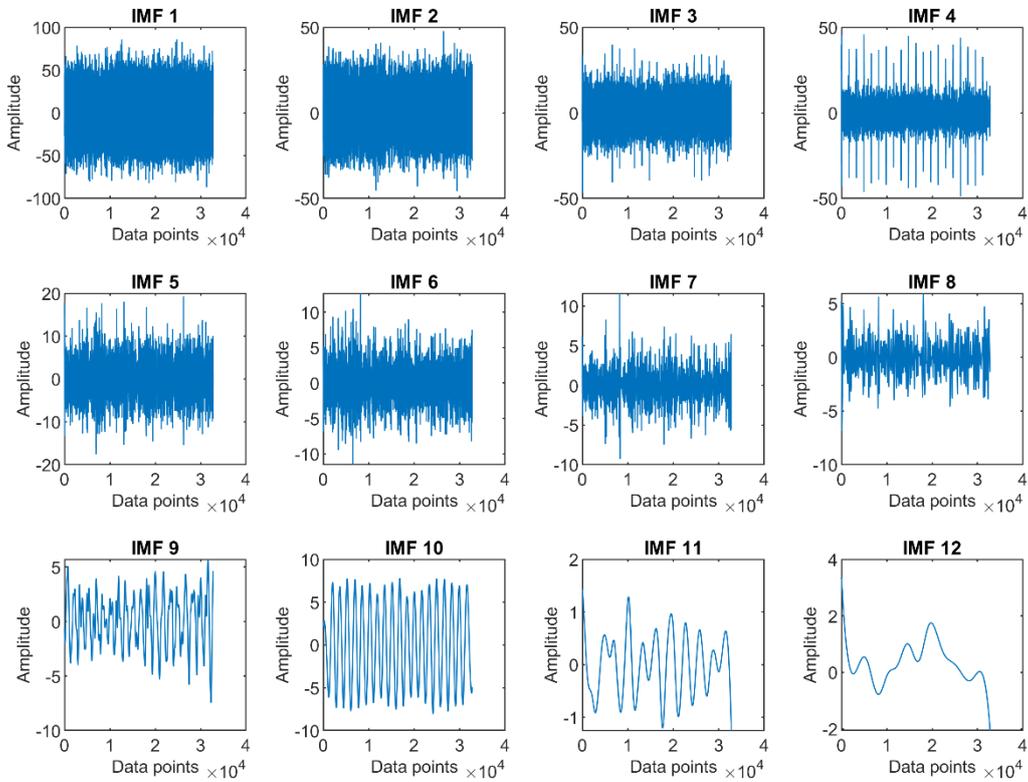

Fig. A17. IMFs generated by applying NPCEEMD to the signal shown in Fig. 8 using ensemble number equal to 100